\documentclass[12pt]{article}
\usepackage{amsmath,amssymb}
\usepackage[dvips]{graphicx}
\usepackage{mathrsfs}
\def\Slash#1{\ooalign{\hfil/\hfil\crcr$#1$}}
\begin{document}
\title{A  new method for measuring   the absolute neutrino
mass }
\author{Kenzo Ishikawa and Yutaka Tobita\thanks{This author contributed equally to this work}}
\maketitle
\begin{center}
Department of Physics, Faculty of Science, \\
Hokkaido University Sapporo 060-0810, Japan
\end{center}

\begin{picture}(0,0)(0,0)
 \put(320,230){\makebox{EPHOU-12-003}}
\end{picture}
\begin{abstract}
{The  probability of the event that  a  neutrino produced in  pion decay
is detected in the intermediate $T$  shorter than the life-time $\tau_{\pi}$, 
$T \leq \tau_{\pi}$,  is sensitive  to the  absolute
 mass of the neutrino.  With a newly formulated  S-matrix $S[T]$ that
 satisfies the boundary conditions of the experiments at a finite $T$, the
 rate of the event  is computed as  $\Gamma_0+\tilde
 g(\omega_{\nu}, {T};\tau_{\pi}) \tilde \Gamma_{1} $, where   $\tilde
 g(\omega_{\nu},{T};\tau_{\pi})$  depends
 weakly on  $\tau_{\pi}$ and  $\omega_{\nu}={m_{\nu}^2c^4}/
{(2E_{\nu}\hbar)}$,  $c$ is the speed of light.  $\Gamma_0$ is 
 the standard one  and the correction,  
$\tilde{g}(\omega_{\nu}, {T};\tau_{\pi}) \tilde \Gamma_{1} $,
reflects relativistic invariance and is rigorously computed  via the  
    light-cone 
singularity of the system   and    
reveals the  
diffraction pattern of a  single  quantum. The formula explains unsolved 
anomalies of neutrino experiments and indicates the heavy  
 neutrino mass,  $0.098 \pm 0.022$ or  $0.083 \pm 0.026$ {eV}/ $c^2$ 
for normal  or  inverted  mass hierarchies, respectively.}
\end{abstract}
\maketitle
\section{Neutrino interference} 
The probability of the event that a particle produced in a decay or
scattering is detected at a finite time-interval $T$ was known to deviate 
from that of 
Fermi's golden rule \cite{peierls}. It was found recently
\cite{Ishikawa-Tobita-Fermi} that  the
probability has a constant term in addition to a $T$-linear term, 
\begin{eqnarray}
P=  \Gamma_{0}  T+P^{(d)},
\end{eqnarray}
where  $\Gamma_{0}$ is the known rate and $P^{(d)} $ is
constant. $P^{(d)} $ has an origin in kinetic-energy non-conserving term  
at a finite $T$, and a new energy scale, ${m^2c^4 \over E}$, where  $m$
and $E$ are the mass  and energy of the detected particle.   $P^{(d)} $ is
proportional to a size of the particle's wave function, and  is smaller 
than $ \Gamma_{0} T$ in many  situations and  has been ignored. As 
regards its relevance with  physical phenomena,  $P^{(d)} $'s magnitude 
is a key. This term is 
  necessary in the processes of
 $\Gamma_{0}=(\approx ) 0$,
  or small $T$. There are two cases in the former,
\begin{eqnarray}
& & \Gamma_0=0,\ P^{(d)}=0, \label{case1}\\
& &\Gamma_0=(\approx) 0,\ P^{(d)}\neq 0. \label{case2}
\end{eqnarray}

Equation $(\ref{case1})$ corresponds to a process forbidden by an exact
symmetry such as charge conservation, and  Equation $(\ref{case2})$
corresponds to a process forbidden by
the  conservation-law of kinetic energy. Equation $(\ref{case1})$ is a trivial
case, but  Equation $(\ref{case2})$ is a non-trivial case.
 $P^{(d)}$ must be included for the study of the 
process connected with this transiton. 
 We study  an example of  Eq. $(\ref{case2})$ that involves  
 light particles.  

Neutrinos  produced in pion
decay reveal  large finite-size corrections  
\cite{Ishikawa-Tobita-Fermi}.   
Neutrino physics  in this region  has   been less explored and all previous
 experiments are not in-consistent with the presence of $P^{(d)}$
 within experimental uncertainties  \cite{pdg}.
In the asymptotic region,  flavor oscillations have been  observed 
and made  to determine the mass-squared difference and other parameters 
possible. The present paper  shows that  the  probability of the event that 
the neutrino is detected at   $T \leq \tau_{\pi}$, where  $\tau_{\pi}$ 
is the pion's life-time,
is  observable  and supplies  
the  information on the neutrino mass.  A new 
experimental method for determining the  absolute neutrino mass is presented.

In the present  region,  kinetic energy of the daughters is not constant, due
to finite interaction energy.
  Thus 
the state   becomes   non-uniform in time and reveals an
interference. The variation of kinetic energy, $\Delta E$,   is  extremely 
small of the order  of $G_F^2$, where $G_F$ is the Fermi coupling
constant, 
and causes the 
interference pattern  to the amplitude and probability 
in $ T \leq  {\hbar \over \Delta E}$. 
The  pattern  formed in the amplitude in a microscopic region initially 
 is  transmitted to large  region, because a group of 
neutrino waves  have  almost the same phase and group velocity $\vec{v}$ 
and can move in parallel for a certain period, $\delta t$, with  relative 
phases kept constant. Thus the   pattern  appears  even at a macroscopic distance  
much larger than    de Broglie wave length. 
Photon,  neutrino, and other light particles or waves  show this 
macroscopic quantum effect. The neutrino  in the leptonic or
semi-leptonic decays  such as $\pi(K) \rightarrow lepton + \nu$ or $ 
K \to \pi+lepton+\nu $ shows this phenomenon but the massive particles 
such as the charged leptons in the above processes or mesons  in the 
non-leptonic decays  such as  $K^{\pm} \rightarrow \pi^{\pm}+\pi^0$ do not.

To compute $P^{(d)}$  involves the consideration of boundary
 conditions. The transition rates at $T$ were computed  before with $S[\infty]$ which
 satisfies the boundary condition at $ T=\infty$, and 
 did not show  clear $T$-dependence 
\cite{khalfin,winter,GW-paper,ekstein,gaemer,peres,maiani-testa}.
Now, the probability of the event that the particle is detected at $T$
must be computed with the amplitude which satisfies the boundary 
conditions at $T$. The correct amplitude   is not
represented with $S[\infty]$.
To compute the probability of the
experiments   at  the  $T \leq \tau_{\pi}$  \cite{GW-paper},  the  S-matrix,  $S[T]$,
defined  by wave functions  
of the initial and final states at  $T$,  which satisfies  the boundary 
conditions at   $T$,  was formulated   in
Ref. \cite{Ishikawa-Tobita-Fermi}.
 The deviations of the rates from those obtained by  Fermi's golden  
rule  were found by using $S[T]$ \cite{Ishikawa-Tobita-Fermi}. 
They  are large for the 
neutrinos and small for the charged leptons.  The corrections depend on 
the boundary
 conditions at $T$, and have the form of $1/{T}$ of   universal properties 
determined by the parameters of Lagrangian. They   become significant 
in  forbidden processes  of light particles in which the rate $\Gamma_0$ vanishes,
and  are  detectable  in  experiments of an energy resolution of much 
larger than $\Delta E$ but of the  order of $(G_F)^0$.   

 We study  the probability in detail and show that  
the finite-size correction to the probability  that depends on   the absolute  
neutrino masses emerges  at a macroscopic ${T}$. Because neutrinos  
interact extremely weakly with matter and are not 
disturbed by environment, the effects are easily observed.
The detailed analysis of the neutrino spectrum and its implications to the  neutrino mass are presented.
 
Comparing   the neutrino spectrum  with previous experiments at $T \leq \tau_{\pi}$, 
  we find  an indication of  the heavy   neutrino mass,   $0.083 \pm 0.026$ for inverted or $0.098 \pm 0.022$ $\text{eV}/ c^2$ for normal mass 
hierarchies, from  its comparison   with LSND.
\section{Wave function of pion and daughters and $S[ T]$}
 $S[T]$ is determined from the wave functions of the in-coming
and out-going waves at $T$ in the system described  with  a Lagrangian
density composed of  a field of pion, $\varphi_\pi(x)$, of 
charged lepton, $l(x)$, and of  neutrino, $\nu(x)$. That has a four-Fermion
coupling     
\begin{align}
&\mathscr{L}=\mathscr{L}_0+\mathscr{L}_\text{int},\label{lagrangian}\\
&\mathscr{L}_0={\partial_{\mu}}\varphi_\pi^{*}{\partial^{\mu}}\varphi_\pi-m_{\pi}^2 \varphi_\pi^{*}\varphi_\pi+\bar{l} (\Slash{p}-m_l)l+\bar\nu(\Slash{p}-m_{\nu})\nu \nonumber, \\
&\mathscr{L}_{int}= g J_\text{hadron}^{V-A}\times J_\text{lepton}^{V-A} +
\mathscr{L}_\text{counter},\ g= G_F/{\sqrt 2}, \nonumber
\end{align}
where 
  $J_i^{V-A}$ is $V-A$
current, and  $\mathscr{L}_\text{counter}$ is a counter term of pion
field. In this paper $\mathscr{L}_\text{counter}=\delta \omega_0
\varphi_\pi^{*}\varphi_\pi $ is applied. Majority of the results are obtained
from the lowest order terms in perturbative expansions, and are the same 
in   electro-weak 
gauge theory.   The case without the flavor mixing  is 
studied first for clarifying the essence easily.
  Due to Poincare invariance,  
momenta of the fields can be any value from $-\infty$ to $+\infty$.  
\subsection{Wave function}
 $|\Psi(t) \rangle$  of the pion  and  decay products  satisfies  the Schr\"{o}dinger equation,
\begin{eqnarray}
i\hbar{\partial \over \partial t}|\Psi(t)\rangle=(H_0+H_\text{int})|\Psi(t) \rangle,
\label{Schro. equation}
\end{eqnarray} 
where $H_0$ is derived from $\mathscr{L}_0$ and  $H_\text{int}=-\int d^3x \mathscr{L}_\text{int}$.
  A time-dependent solution in the 
first order of $H_\text{int}$ of the initial condition at $t=0$
\begin{eqnarray}
|\psi^{(0)} \rangle=|{\vec p}_{\pi} \rangle
\end{eqnarray}
is
\begin{align}
&|\Psi(t) \rangle=e^{(-i\frac{E_0}{\hbar}-{1 \over \tau_{\pi}})t}|\psi^{(0)} \rangle +e^{(-i\frac{E_0}{\hbar})t}\int
 d{\beta}D(\omega,t) 
|\beta \rangle \langle \beta|H_\text{int}|\psi^{(0)} \rangle
 \label{pertubative-wave},\\
&\omega =E_{\beta}-E_0, \ H_0|\beta \rangle=E_{\beta}|\beta \rangle,\ H_0|\psi^{(0)}\rangle=E_{0}|\psi^{(0)}\rangle,\nonumber\\
&D(\omega,t)=\frac{e^{-i\frac{\omega }{\hbar}t }-e^{-{t \over \tau_{\pi}}}}{\omega+i{\hbar \over 
\tau_{\pi} }}.\nonumber
\end{align} 
In Eq. $(\ref{pertubative-wave})$, the pion's life-time, $\tau_{\pi}$, given as an imaginary part 
of the second  order correction 
\begin{eqnarray}
\tau_\pi={\hbar \over \Gamma_0}, ~~\Gamma_0=\frac{G_F^2f_{\pi}^2}{8\pi}
m_{\pi}^2m_{l}^2\left(1-{m_{l}^2 \over m_{\pi}^2}\right)^2{1 \over E_{\pi}} \label{life-time},
\end{eqnarray}
where $f_{\pi}$ defined by  
\begin{eqnarray}
\label{pion-coupling}
\langle 0 |J_{V-A}^{\mu}(0)|\pi \rangle =if_{\pi} p_{\pi}^{\mu},
\end{eqnarray}
  was included, and $d\beta$ is a measure for a complete set of 
$|\beta \rangle=|\nu,l \rangle$. For the
real part of the pion energy, the divergences are subtracted by  
the counter term which expresses 
the renormalization of field operator 
and mass.

 At a finite $t$,  $|\Psi(t) \rangle$ is a superposition of the parent   and daughters and has 
a  finite interaction energy.
 Consequently the kinetic energy deviates
from that of the initial energy, and varies.
  The energy difference 
\begin{eqnarray}
\Delta E(t)=\langle \Psi (t)|\left(i\hbar {\partial \over \partial t}-E_0\right)|\Psi(t) \rangle/
N 
\end{eqnarray}
shows  an order-parameter of expressing the wave nature.  $\Delta E(t) $ 
 vanishes  in free particles, and is finite  in  waves due to the interference. 
   $N_0=(2\pi)^3\delta^{(3)}(0)=V$, where $V$ is a normalization volume,  is used to factor out the normalization of the state.
Substituting Eq. $(\ref{pertubative-wave})$, we have 
\begin{align}
\Delta E(t)&=e^{-\frac{t}{\tau_{\pi}}}\int d\beta [D^{*}(\omega,t)+D(\omega,t)]|\langle \psi^{(0)}|H_{int}|\beta \rangle|^2 \nonumber\\
&=e^{-\frac{t}{\tau_{\pi}}}A_1+e^{-\frac{2 t }{ \tau_{\pi}}}A_2,
\label{energy-deviation}
\end{align}
where 
 $A_1$ and $A_2$ are 
\begin{align}
&A_1= {G_{F}^2 f_{\pi}^2 \over 2\pi^2}m_{\pi}^2m_l^2 \int_{-\infty}^{m_{\pi}-m_l} d \omega (1-{m_l^2 \over (m_{\pi}-\omega)^2})^2{\omega \cos \omega t/{\hbar} +{\hbar /\tau_{\pi}\sin \omega t/{\hbar}} \over \omega^2+({\hbar} /{\tau_{\pi}})^2},\\
&A_2=  {G_{F}^2 f_{\pi}^2 \over 2\pi^2} m_{\pi}^2m_l^2 \int_{-\infty}^{m_{\pi}-m_l} 
d \omega {m_l^2 \over (m_{\pi}-\omega)^2}\left(2-{m_l^2 \over (m_{\pi}-\omega)^2}\right)
{2\omega   \over \omega^2+({\hbar} /{\tau_{\pi}})^2}, 
\end{align} 
with a choice of  $\delta \omega^0$, although this is not unique, as 
\begin{eqnarray}
\delta \omega_0 =  {G_{F}^2 f_{\pi}^2 \over 2\pi^2}m_{\pi}^2m_l^2 \int_{-\infty}^{m_{\pi}-m_l} 
 \frac{2 \omega d \omega }{\omega^2+({\hbar} /{\tau_{\pi}})^2}. 
\end{eqnarray}
$\Delta E(t)$ for the normalization $N=\Gamma N_0$ is roughly the energy 
difference per event.

At  $t=\infty$,  $\Delta E(\infty)=0$, and  the 
first term in $|\Psi(t) \rangle$ vanishes  and the state agrees with 
\begin{eqnarray}
& &|\Psi_{\infty} \rangle= -
 2\pi ie^{-i\frac{E_0}{\hbar}t}\int d{\beta} e^{-i\omega {\hbar}t}{1 \over \omega+i\hbar/{\tau_{\pi}}} |\beta \rangle \langle \beta
 |H_{int}|\psi^{(0)} \rangle , \label{infinite-time}  \\
& &H|\Psi_{\infty} \rangle=E_0|\Psi_{\infty} \rangle,\ H_0|\Psi_{\infty}
 \rangle=E_0|\Psi_{\infty} \rangle.  \nonumber
\end{eqnarray}
 The asymptotic state is composed of free particles  of  the total  and  
kinetic energy of the region $E_0\pm{\hbar \over \tau_{\pi}}$.

From Eq. \eqref{energy-deviation},   $\Delta E(t) \neq 0$, and
 the state   Eq. $(\ref{pertubative-wave})$ is the  sum of
$|\psi^{(0)}\rangle$ and $|\beta \rangle$ of  continuous 
  $E_{\beta} \geq 0$ in $t\leq \tau_{\pi}$. The state of varying
 kinetic-energy is wave-like, and  reveals  non-uniform probability in $t$. 
 The probability of  event that the   daughters are detected 
at finite-time interval $T$ depends on $T$. 
\subsection{$ S[ T]$}
Physical quantities are measured through  transition processes.  A
transition  amplitude at a finite $T$  is 
uniquely defined with a set of 
initial and  final states,  and   a time 
interval, and  is represented by the  S-matrix that
satisfy the boundary condition at $T$, i.e., $S[T]$. 
 $S[T]$ holds various unusual properties 
which   are different form $S[\infty]$ and has  been barely discussed in the literature.  $S[T]$  is defined in Heisenberg 
representation by the boundary conditions for   
field operators \cite{Ishikawa-Tobita-Fermi},  as an extension of the standard
LSZ formalism \cite{LSZ,Low}.
For a scattering 
from an initial state 
 $|\alpha \rangle $ at $t=-T/2$ to a final state
$|\beta \rangle$ at $t=T/2$ of a scalar field expressed by 
$\varphi(x)$, where   
$|\alpha \rangle$ are
constructed with free 
waves $\varphi_\text{in}(x)$ and 
$|\beta \rangle $ at $t=T/2$ are constructed with free waves
$\varphi_\text{out}(x)$,   boundary
conditions are 
\begin{eqnarray}
& &\lim_{t \to -{T}/2}\langle \alpha |\varphi^{f}(t)|\beta
 \rangle= \langle \alpha|\varphi^{f}_\text{in}|\beta \rangle,
\label{boundary-condition1}\\
& &\lim_{t \to +{T}/2}\langle \alpha |\varphi^{f}(t)|\beta
 \rangle=  \langle \alpha |\varphi^{f}_\text{out}|\beta \rangle,
\label{boundary-condition2}
\end{eqnarray}   
where  $\varphi_\text{in}(x)$ and $\varphi_\text{out}(x)$ satisfy the free wave
 equation \footnote[1]{$Z^{1/2}$,  multiplied in the right-hand
 sides of the above equations are 1 in the present order.}. 
  $\varphi^{f}(t)$  is  the expansion coefficient of field $\varphi(x)$ 
with c-number function $f(x)$ as
\begin{eqnarray}
\varphi^{f}(t)=i \int d^3 x
f^{*}({\vec x},t)\overleftrightarrow{\partial_0} \varphi({\vec x},t).
\label{basis-expansion}
\end{eqnarray}
$\varphi_\text{in}^{f}$ and $\varphi_\text{out}^{f}$ are defined in the same way.
The function  $f({\vec x},t)$  is a normalized solution of   free wave
equation and    decreases fast   at 
large $|{\vec x}-{\vec x}_0|$  around the center ${\vec x}_0$, and is
denoted as a wave packet.  
Thus the states $|\alpha  \rangle$ and   $|\beta  \rangle$, and    
the boundary conditions Eqs. $(\ref{boundary-condition1})$ and
 $(\ref{boundary-condition2})$ depend  on 
 wave packets.

  $S[T]$  is expressed   by  
M{\o}ller operators of the 
finite-time
interval, $\Omega_{\pm}({T})$, as $S[T]=\Omega_{-}^{\dagger}(T)\Omega_{+}(T)$,
 and satisfies 
\begin{align}
&\left[S[T],H_0\right]=  i\left\{\frac{\partial}{\partial
	{T}}\Omega^{\dagger}_{-}({T})\right\}\Omega_{+}({T})-
	i\Omega^{\dagger}_{-}({T}){\partial \over \partial {T}}
	\Omega_{+}({T}),
\label{commutation-relation}
\end{align}
where 
\begin{eqnarray}
\Omega_{\pm}(T)=\displaystyle\lim_{t
 \rightarrow \mp T/2}e^{iHt}e^{-iH_0t}.
\end{eqnarray}
Hence a   matrix
element of $S[T]$ between a state $|\alpha \rangle$ 
and another state $|\beta \rangle $ 
is written as the sum of the energy-conserving and non-conserving terms,
\begin{eqnarray}
\langle \beta |S[{T} ]| \alpha \rangle=\langle \beta
 |S^{(n)}[{T} ]| \alpha \rangle+ \langle \beta |S^{(d)}[{T}]|
 \alpha \rangle.
\label{matrix-element}
\end{eqnarray}
Expanding  $|\alpha \rangle$ and $|\beta \rangle$  with  eigenstates  
of $H_0$ of eigenvalue  $E_{\alpha}$ and $E_{\beta}$, we have
\begin{eqnarray}
\sum_{E_{\beta}} \left[\langle \beta|E_{\beta} \rangle \langle E_{\beta}
 |S^{(n)}[{T}]| E_{\alpha} \rangle \langle E_{\alpha} |\alpha  \rangle + \langle \beta |E_{\beta} \rangle
 \langle E_{\beta}|S^{(d)}[{T}]|E_{\alpha}  \rangle \langle E_{\alpha}| \alpha  \rangle \right],
\label{matrix-element2}
\end{eqnarray}
where $E_{\beta}=E_{\alpha}$ in $S^{(n)}[T]$  and 
$E_{\beta} \neq E_{\alpha}$ in $S^{(d)}[T]$. Thus  the finite number of
states  couple with $S^{(n)}[T]$, and infinite number of 
states   could couple with  $S^{(d)}[ T]$. Among those states  in the
latter,  
the states  determined by the
boundary conditions Eqs. $(\ref{boundary-condition1})$ and
 $(\ref{boundary-condition2})$ couple. They necessary depend on $f(x)$ from
Eq. ($\ref{basis-expansion})$, and  $S^{(d)}[T]$ depends on $f(x)$ and is
appropriate to write as $S^{(d)}[ T;f]$. 
The right-hand side
of Eq. $(\ref{commutation-relation})$ and   $S^{(d)}[ T;f]$  vanish 
at $T \to \infty$, and 
 are finite  at a finite $T$.
Thus  $S^{(d)}[ T;f]$     gives 
the finite-size correction.  
Because $|E_{\beta} \rangle$ and $|E_{\alpha} \rangle $ are orthogonal if $E_{\beta} \neq E_{\alpha}$,
  the cross 
term  in a square of the modulus  of the  first  and second terms of
Eqs. $(\ref{matrix-element})$ and $(\ref{matrix-element2})$ vanish, and  the  finite-size 
correction  becomes   positive
semi-definite and increases with the number of states.

$\langle E_{\beta}|S^{(d)}[T;f]|E_{\alpha} \rangle$ 
is proportional to $G_F$ from Eq. $(\ref{commutation-relation})$, so is 
$\langle E_{\beta}|S^{(n)}[T]|E_{\alpha}
\rangle$. Since
\begin{eqnarray} 
(E_{\beta}-E_{\alpha}) \langle E_{\beta} |S^{(d)}[T;f]|E_{\alpha} \rangle =\langle
 E_{\beta} |[S[\text T],H_0]|E_{\alpha} \rangle,
\end{eqnarray}
$E_{\beta} - E_{\alpha}$ can be as large as $(G_F)^0=1$.


  $f({\vec x},t)$ in Eqs. $(\ref{boundary-condition1})$
 and $(\ref{boundary-condition2})$
 expresses  the wave functions of microscopic objects involved   in
 the processes.  They are extended in large space for propagating waves 
 but are localized  in small space for bound states in matter.   The pion, 
charged lepton, and neutrino behave differently.

A pion is in the initial state in the decay process. 
A high-energy pion produced in proton collisions in matter has a large
  mean free
 path and is approximately
 represented by  a
plane-wave  \cite{Ishikawa-Shimomura,Ishikawa-Tobita-ptp,Ishikawa-Tobita,Ishikawa-Tobita1}.
A nucleon in a nucleus in solid  has a  small size. However  a proton in
  a beam of high energy experiments propagates almost freely and is 
expressed with  wave  function of large size. In their collisions, they 
overlap for a time interval determined by the latter size  hence  the 
produced pion has this large size. If that had a size of nucleus, the decay 
products from  this   pion of nucleon size would have had such
large  energy spreading that  is in-consistent with experiments as given 
in Ref. \cite{Ishikawa-Tobita-Fermi}. From these reasons, the large  
wave packet is suitable for the pion in the initial state 
\footnote[2]{Actually low energy negative 
pion and muon bound in matter may be described by small wave packets, and will 
be studied in a separate publication.}.
 Arguments for  the small size  of the order of a nucleon   
in matter was given in Ref. \cite{Akhmedov2},  but the neutrinos
  produced from the decay of 
  pion of small wave functions have the small sizes and are separated
  easily. They do not show    
flavor  oscillation observed in  the long baseline experiments. Hence
  the small wave packets for the pion is not appropriate. 

A charged lepton  is in the final  state in the decay process. 
A charged lepton produced in the pion decay is  undetected in
 the neutrino experiments and  
 can be  expressed with any functions. Here the simplest  plane wave is
 used. 
The neutrino  is detected  indirectly by observing 
particles produced by  incoherent collisions of the neutrino  with
 nucleus  in target.    The 
nucleus  is expressed by a localized function, and 
momenta of reaction products  
are measured within certain uncertainties determined by the nucleus size 
around  $10$ -- $100$ MeV/$c$. 

A neutrino interacts with matter extremely weakly and has a large mean
free path, of the order of $10^9$ m for $E_{\nu}=1$ GeV in the earth. Hence for a process of
in-coming neutrino, $ \sigma_{\nu} =\infty$ obtained from  the mean free
path is used. For a process
of out-going neutrino, $\sigma_{\nu}$ becomes completely different from
the above value due to boundary conditions. 
In the amplitude of the events that the neutrino is 
detected by or  interacts with  a nucleus in solid at a position,  
${\vec X}$, the neutrino is expressed by the nucleus wave function. The 
nucleus   
is a bound state and expressed with a normalized small wave function.
  $S[ T]$
thus constructed satisfy the boundary condition of the experiment, and
is the correct one. $ \sigma_{\nu} =\infty$ is suitable for the
in-coming state but the size of nucleus, which is  small but 
non-zero, is suitable for the out-going state. The position ${\vec X}$
is not identified and the probability added over    ${\vec X}$ is measured.

In a treatment of the whole process where  the neutrino is not the out-going
state but expressed with a propagator  as  an 
intermediate state,   in-coming states has  many nucleus and a
pion. Nucleus in solid are located in distance positions each others 
and are treated incoherently.  This amplitude  for a large-time
interval, is written as  a 
 product of  the amplitude of the pion, lepton, and an on-mass shell
 neutrino, at a coordinate x, and that of the neutrino, nucleus, and a 
final state that  includes a nucleus and  lepton at a coordinate y. The 
nucleus wave functions have  small sizes, and an integration over y is
made easily  around $({\vec X},T_{\nu})$. Finally the former 
part becomes equivalent to the element of $S[ T]$ of 
$\sigma_{\nu}=\sigma_{nucleus}$, and the latter
part becomes the amplitude of the neutrino nucleus collision. 
 Thus in  $S[ T]$,  the neutrino is expressed with  the nucleus 
wave function.
Furthermore, the reaction of the neutrino with the target nucleus 
 is analyzed in fact  by 
Monte Carlo codes using   relativistic  Fermi gas 
  model \cite{fermi-gas}, where a  nucleon is moving freely subject to a 
nuclear potential in the nucleus. 

Thus the out-going neutrino in $S[ T]$ of the event that the
neutrino is detected in pion decay has a small  nucleus size 
despite of the fact that the neutrino propagates freely and has the
infinite mean free path
    \cite{Ishikawa-Shimomura,Ishikawa-Tobita-ptp,Ishikawa-Tobita,
  Ishikawa-Tobita1}.

   A complete set of small wave packets includes the  
center position in addition to the momentum \cite{Ishikawa-Shimomura}.
Matrix elements of $S[T]$ and the probability depend 
 on  the  center position in addition to the momentum.
 $S[\infty]$ gives the probability  at $T=\infty$ and the total 
probability is independent of the functions $f({\vec x},t)$.   Previous
 works on wave packets have taken the boundary condition at $T=\infty$, 
accordingly,  they have computed the asymptotic values.  In 
$T \gg \tau_{\pi}$,  they show 
flavor oscillations, which    depend on the mass-squared differences 
$\delta m_i^2;\ i=1-3$ and  agree with the standard formula  based on  single particle picture
\cite{Kayser,Giunti,Nussinov,Kiers,Stodolsky,Lipkin,Akhmedov,Asahara}
 or in field theory
\cite{Kayser,Giunti,Nussinov,Kiers,beuthe,Akhmedov,Asahara,Akhmedov2}.  The pion's life-time was studied within 
the framework of $S[\infty]$ and shown not to 
modify the formula   \cite{Grimus}.
  Textbooks on  scattering
or decay processes   \cite {LSZ,Low,Goldberger,newton,taylor}, 
and field theory \cite{qft-texts1,qft-texts2} 
 which emphasize the  importance of wave packets and  prove  that 
the asymptotic values with wave packets are equivalent to those of plane waves,
 studied  also $S[\infty ]$ expressed with large wave packets in 
Poincare invariant manners. Accordingly the finite-size corrections
have not been studied in previous works on wave packets. Some signal of 
the correction were found in Ref. \cite{Asahara}, which used the
boundary condition at T.
$S[T]$ and the probability depend on the wave packets from
Eqs. $(\ref{boundary-condition1})$ and  $(\ref{boundary-condition2})$.

\begin{figure}[t]
 \centering{\includegraphics[scale=.40]{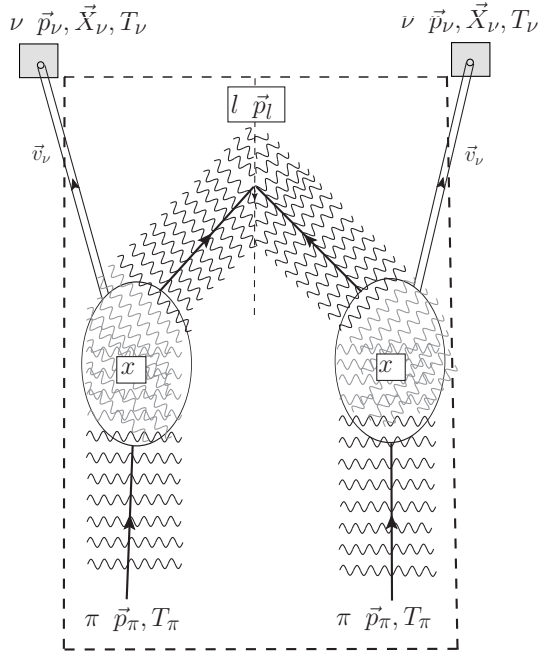}%
  \caption{A space-time view of the product of the  amplitude of the event 
that a neutrino is detected   in a decay of 
a pion    Eq. $(\ref{probability-correlation})$ is given. The pion and lepton 
are expressed by plane waves and the neutrino is by the wave packet of the size $\sigma_{\nu}$.  The  integrand is the product of the neutrino wave function 
of a velocity ${\vec v}_{\nu}$ and  
$\Delta_{\pi,l}(x_1,x_2)$ that corresponds to the
  diagram   surrounded  by a box of dot line.}
  \label{fig:1}}
 \end{figure}


$H_0+H_\text{int}$ in Eq. $(\ref{Schro. equation})$ is Hermitian, and the 
norm of the sate is preserved and  $S[T]S^{\dagger}[T] =1$. 
This unitarity ensures the conservation of probability, i.e.,  a sum 
of the probability  for finding the pion  and  the lepton and neutrino 
at a $t$ simultaneously.  Furthermore, the charged lepton and corresponding 
neutrino have the same probability in this case. If the probabilities are  
measured independently of  particles at $T$, there is no reason for them to be 
the same. The unitarity also does not connect  the probability at $T$,  
$P(T)$, with that at $T'$, $P({T}')$. Consequently the decay rates could depend on 
both of $T$ and measured particles, and are computed with 
$S[T]$ of fulfilling  the boundary conditions of experiments.

\section{Position-dependent probability}  
 We now study  the  probability  of event  that a neutrino is
 detected at a finite distance.  The 
transition amplitude  is 
the scalar product of the  initial wave function   with the 
   final wave function that has  the kinetic energy varying with time. 
The $S[T]$ and  probability reflect this property and show  
non-uniform behavior, which  is not computable correctly with  $S[\infty]$.   
   Details of the derivations 
  are explained in 
Refs. \cite{Ishikawa-Tobita1,Ishikawa-Tobita2}. 
\subsection{High-energy pion}
The decay of a plane wave  of pion to a plane wave of lepton and a wave packet
of neutrino is expressed with a matrix element  of   $S[T]$ between  a 
state composed of the   neutrino of  ${\vec p}_{\nu}$ at $\vec{{X}}_{\nu}$ 
and the  charged lepton $l$ of  ${\vec p}_{l}$  and  the  initial pion  
at $t={T}_{\pi}$ of  ${\vec p}_{\pi}$, Fig. \ref{fig:1}, 
is  expressed as
 $\mathcal{M}=\int d^4x \, \langle {l},{\nu}
 |H_\text{int}(x)| \pi \rangle$,  where   
$
|\pi \rangle=   | {\vec p}_{\pi},{T}_{\pi}  \rangle,\ 
|l ,\nu \rangle=   |\vec{p}_{l};\vec{p}_{\nu},\vec{{X}}_{\nu},{T}_{\nu}          \rangle$.
 In  high energy, the life-time of the pion becomes long and  can be ignored.
$\mathcal{M}$ is then
written 
 with the  matrix element of $V-A$ current of pion, Eq. $(\ref{pion-coupling})$, 
and  Dirac spinors 
\begin{align}
\label{amplitude}
&\mathcal{M} = \int d^4xN_1\langle 0 |J_{V-A}^{\mu}(0)|\pi \rangle 
\bar{u}({\vec p}_l)\gamma_{\mu} (1 - \gamma_5)
\nu(x,{\vec p}_{\nu},{\vec X}_{\nu},T_{\nu})
\exp\left[-i(p_{\pi}-p_l) \cdot x /\hbar\right] , \\
&\nu(x,{\vec p}_{\nu},{\vec
 X}_{\nu},T_{\nu})=\left({\frac{\sigma_{\nu}}{\pi}}\right)^{\frac{3}{4}} \int d{\vec k}_{\nu} \sqrt{ \frac{m_{\nu}}{E_{\nu}}}
 \exp\left[{i{k_\nu\cdot(x - {X}_\nu)/\hbar}-
\frac{\sigma_{\nu}}{2}({\vec k}_{\nu}-{\vec p}_{\nu})^2}\right] \nu({\vec k}_{\nu})\nonumber,
\end{align}
where 
$N_1=ig \left({m_l }/{
 E_l }\right)^{\frac{1}{2}} ((2\pi)^32E_{\pi}V)^{-\frac{1}{2}}$ and $V$ is a normalization volume.
The wave packet $\nu(x,{\vec p}_{\nu},{\vec X}_{\nu},T_{\nu})$ satisfies the free wave equation 
and decreases rapidly at $t=T_{\nu}$ with  ${\vec x}-{\vec X}$  and  satisfies  the  condition for
 $f({\vec{x}},t)$ in Eq. $(\ref{basis-expansion})$ \footnote[3]{If the integration over $x$ is 
made first, $\mathcal{M}$  satisfies the boundary condition  of 
$S[\infty]$, instead of  $S[T]$.}.    
For  $t \leq T_{\nu}$,
\begin{align}
&\nu(x,{\vec p}_{\nu},{\vec X}_{\nu},T_{\nu})
= \left({4\pi \over \sigma_{\nu}}\right)^{\frac{3}{4}}\sqrt{\frac{m_{\nu}}{E_{\nu}(p_{\nu})}}
e^{ip_\nu\cdot(x-X_\nu)/\hbar
-\frac{1}{2\sigma_{\nu}} ({\vec x}-{\vec X}_{\nu}-{\vec v}_{\nu}(t-T_{\nu}))^2} 
\left\{1+O({p_{\nu}^{-1}})\right\},
\end{align}
and  the center  moves   with the velocity ${\vec
v}_{\nu}={\vec p}_{\nu}c^2/{E_{\nu}}$, $\vec{x}_{\,0} =
\vec{{X}}_{\nu} 
+ {\vec v}_\nu(t-{T}_{\nu})$. 
We ignore $O({p_{\nu}^{-1}})$ term. The integrand of Eq. $(\ref{amplitude})$ 
becomes finite along a  narrow space-time region of 
the velocity ${\vec v}_{\nu}$, and  $t$ is
 integrated over  ${T}_{\pi} \leq t \leq {T}_{\nu}$. 
${\sigma_{\nu}}$ is the size
of the neutrino wave packet \cite{Ishikawa-Tobita-ptp,Ishikawa-Tobita, Ishikawa-Tobita1}. 
For the sake of simplicity, we use the Gaussian form of the wave packet
in this paper.  
The  result for   the finite-size correction is the same in general 
wave packets.

The  total probability  is an integral of 
a square of the modulus  of the amplitude over the complete set of final states
\cite{Ishikawa-Shimomura},
\begin{eqnarray}
P=\int d\vec{{X}}_{\nu} {d{\vec p}_{\nu} \over (2\pi)^3} 
  \frac{d{\vec
 p}_l}{(2\pi)^3} \sum_{s_1,s_2}|\mathcal{M}|^2,
\label{total-probability}
\end{eqnarray}
 where the momenta of the neutrino and  charged lepton are integrated over the
whole positive energy region, and the position of the wave packet is
integrated  over the region of the detector. This    depends on
${T}={T}_{\nu}-{T}_{\pi}$. Hereafter the natural unit,
 $c=\hbar=1$, is taken in majority of places, but $c$ and $\hbar$ are
written explicitly when it is necessary. 
Figure \ref{fig:1} shows the space-time configuration 
of the amplitude  and 
probability expressed by the 
overlap of wave functions of the initial pion with those of  lepton and 
neutrino.

Computation of the probability  in a consistent manner
with the Lorentz invariance was made  with
 a correlation function $\Delta_{\pi,l}(x_1,x_2)$ of the pion and lepton 
vertex  and the neutrino wave function.
 Integrating  over the momentum ${\vec p}_l$  first, 
 after summing over the spin,  and we have  the   probability  in the form 
 \begin{align}
P&=\int d\vec{{X}}_{\nu} {d{\vec p}_{\nu} \over (2\pi)^3}  \frac{N_2}{E_\nu}\int d^4x_1 d^4x_2 
\exp{\left[{-\frac{1}{2\sigma_\nu}\sum_i ({\vec
  x}_i-\vec{x}_i^{\,0})^2}+i\phi(\delta x)\right]} \Delta_{\pi,l}(\delta x),
\label{probability-correlation}  \\
\vec{x}_i^{\,0} &= \vec{{X}}_{\nu} + {\vec
v}_\nu(t_i-{T}_{\nu}),\ \delta x
=x_1-x_2,\ \phi(\delta x)=p_{\nu}\!\cdot\!\delta x,\nonumber
\end{align}
where $N_2=g^2f_{\pi}^2
\left({4\pi}/{\sigma_{\nu}}\right)^{\frac{3}{2}}((2\pi)^32E_{\pi}V)^{-1}$,   and 
\begin{eqnarray}
\Delta_{\pi,l} (\delta x)=
 {\frac{1}{(2\pi)^3}}\int
{d {\vec p}_l \over E({\vec p}_l)}\left\{2(p_{\pi}\cdot p_{\nu})( p_{\pi}\cdot
 p_l)-m_{\pi}^2 (p_l \cdot p_{\nu})\right\}e^{-i(p_{\pi}-p_l)\cdot\delta x }.
\label{pi-mucorrelation}
\end{eqnarray}

$\Delta_{\pi,l} (\delta x)$ becomes   the sum of  
the light-cone singularity, $\delta({\delta
x}^2)$,  and less singular and regular functions in the region
 $m_{\pi}^2 \geq m_l^2$. $\Delta_{\pi,l} (\delta
 x)$ vanishes in  $m_{\pi}^2 < m_l^2$.
 $\Delta_{\pi,l}(\delta x)$ is then expressed as
\begin{align}
\Delta_{\pi,l}(\delta x)=&2i
\left\{m_{\pi}^2p_{\nu}\cdot\left(p_{\pi}+i \frac{\partial}
 {\partial \delta x} \right) -2i(p_{\pi}\cdot p_{\nu})
 \left(p_{\pi}\cdot\frac{\partial}{
 \partial \delta x}\right)\right\}
\nonumber\\
&\times\left[\frac{\epsilon(\delta t)}{4\pi}\delta(\lambda)+
 I_1^\text{regular}
+I_2\right],
\label{muon-correlation-total}
\end{align}
where $\lambda=(\delta x)^2 = \delta t^2 - \delta\vec{x}^{\,2}
$ and $I_1^\text{regular}$ is composed of  Bessel functions (see Appendix B).
$\epsilon(\delta t)$ is a sign
function and $\delta(\lambda)$ is Dirac's delta function. $I_2$ is regular.
After tedious integrations over ${\vec x}_i(i=1,2)$ in Eq. $(\ref{probability-correlation})$, we have the slowly 
varying term  from the 
light-cone singularity  
\begin{align}
&J_{\delta(\lambda)}=C_{\delta(\lambda)}
\frac{\epsilon(\delta t)}{|\delta t|
 }\exp\left[{i\bar \phi_c(\delta t)-\frac{m_{\nu}^4c^8}{
 16\sigma_{\nu} E_{\nu}^4} {\delta
 t}^2}\right],\label{lightcone-integration2-2} \\
&C_{\delta(\lambda)}=\frac{{(\sigma_{\nu}\pi)}^{\frac{3}{2}}
 {\sigma_{\nu}}}{2},\ \bar\phi_c(\delta t)=\omega_{\nu} \delta t=\frac{m_{\nu}^2c^4}{
 2E_{\nu}}\delta t,\nonumber
\end{align} 
and rapidly decreasing or oscillating functions from $I_1^\text{regular}$.
 The phase $\phi(\delta x)$ that varies rapidly in $\delta t$ and $\delta {\vec x}$
 in Eq. $(\ref{probability-correlation})$ became $\bar \phi_c(\delta t)$ of 
 the slow angular velocity $\omega_{\nu}$ in 
$\bar \phi_c(\delta t)$ of Eq. $(\ref{lightcone-integration2-2})$ at the
 light cone $\lambda=0$.  
The next singular 
term is from  
${1/\lambda}$ in $\Delta_{\pi,l}(\delta x)$, and becomes  
$ J_{\delta(\lambda)} / \sqrt {\pi \sigma_{\nu}
 |\vec{p}_{\nu}|^2}$. This is much smaller than $ J_{\delta(\lambda)}$ and is
 negligible in the
 present parameter region.
 The
 magnitude is inversely proportional to  $|\delta t|$
. This behavior is satisfied in 
general forms of the wave packets.


Finally  we have the probability  in the form, 
\begin{align}
\label{total-probability2}
&P=\int d\vec{{X}}_\nu {d^3p_{\nu} \over (2\pi)^3}N_3\int dt_1 dt_2\biggl[    
\frac{\epsilon(\delta t)}{|\delta t|}e^{i {\bar \phi_c(\delta t)}}
 +2D_{\tilde m}(p_{\nu}){\tilde L_1 \over \sigma_{\nu}}-{2i \over
 \pi}\left( {\sigma_\nu \over \pi}\right)^{\frac{1}{2}}\tilde
 L_2\biggr], \nonumber\\
&N_3=i{\sigma_{\nu}} g^2 f_{\pi}^2 p_{\pi}\!
 \cdot\! p_{\nu}(m_{\pi}^2-2p_{\pi}p_{\nu})(2E_{\pi}E_{\nu} V)^{-1},
\end{align}
where $D_{\tilde m}(p_{\nu})$ and $\tilde L_2$
 are given in Appendix B.
 The first term in Eq. (\ref{total-probability2}) oscillates extremely slowly  
with the angular velocity $\omega_{\nu}$. The remaining terms tend 
exponentially to  zero  or oscillating functions, as $|\delta t|
 \rightarrow \infty$.  The first and second terms in the right-hand side of 
Eq. $(\ref{total-probability2})$ exist   
in the 
region
\begin{eqnarray}
2p_{\pi}\! \cdot\! p_{\nu}\leq {\tilde m}^2 =m_{\pi}^2-m_l^2,
\label{convergence}
\end{eqnarray}
 and vanish outside  this region \cite{Ishikawa-Tobita-Fermi}.

Integrations over $t_1$ and $t_2$ are made next. The slowly varying and rapidly varying terms, $ \frac{1}{\delta t }e^{i\omega_{\nu} \delta t} $ and  $
 I_1^\text{regular}$,  give the probabilities of different behaviors. 
The integrations over  $t_1$ and $t_2$  of the former  term  is 
\begin{align}
&i  \int_0^{{T}} dt_1 dt_2  \frac{\epsilon(\delta t)}{|\delta t|}e^{i {\omega_{\nu}}\delta t }   
= {T} (\tilde g(\omega_{\nu}{T})-\pi),\ 
\label{probability1} 
\end{align}
where the function $\tilde
g(\omega_{\nu}{T})$ in the right-hand side satisfies
$\tilde{g}(\omega_{\nu}{T})_{{T}=0}=\pi,\frac{\partial}{\partial 
{T}}\tilde g(\omega_{\nu}{T})|_{{T}=0}
 = -\omega_{\nu}$ and  $\tilde{g}(\omega_{\nu}T)=\frac{2}{
 \omega_{\nu} T},\text{ for } \omega_{\nu}T \rightarrow \infty$. 
The last term in Eq. ($\ref{probability1}$) is canceled by 
the integral of the short-range term $I_1^\text{regular}$. 
 $\tilde{g}(\omega_\nu{T}) $ is generated by
the sum of   infinite waves that  results to the light-cone singularity, and we call 
this a diffraction term.
The  last  term $I_2$ in
Eq. $(\ref{muon-correlation-total})$  gives   $
{2 \over \pi}\sqrt{\sigma_\nu \over \pi}\int dt_1 dt_2 \tilde
 L_2(\delta t) ={T} G_0$,
where the constant $G_0$ is computed numerically.
 Owing  to the rapid oscillation, 
only the microscopic $|\delta t|$ contributes to this integral, and
consequently  $G_0$ is constant in a macroscopic $T$.

Integrating over $\vec{{X}}_{\nu}$,
we obtain the total
probability  expressed as the sum of the normal term $G_0$ and the
diffraction term $\tilde g(\omega_\nu{T})$:
\begin{align}
\label{probability-3}
P=N_4\int \frac{d\vec{p}_{\nu}}{(2\pi)^3}
\frac{p_{\pi}\! \cdot\! p_{\nu}(m_{\pi}^2-2p_{\pi}\! \cdot\! p_{\nu}) }{E_\nu}
 \left[\tilde g(\omega_{\nu}{T}) 
 +G_0 \right],
\end{align}
where $N_4 = {T}g^2 f_{\pi}^2\sigma_\nu(2E_{\pi})^{-1}$ and ${L} = c{T}$ is the
length of the decay region.  $P$ is the probability of the event that the neutrino is detected. The first term in the right-hand side  corresponds to the
energy-non-conserving term $S^{(d)}[{T};f ]$, and the second  term 
corresponds to the energy-conserving term $S^{(n)}[{T}]$. 
The former vanishes at ${T} \rightarrow \infty$, and is the
finite-size correction, which is stable with respect to variation of the
pion's momentum.        
 \begin{figure}[t]%
  \begin{center}
   \includegraphics[angle=-90,scale=.40]{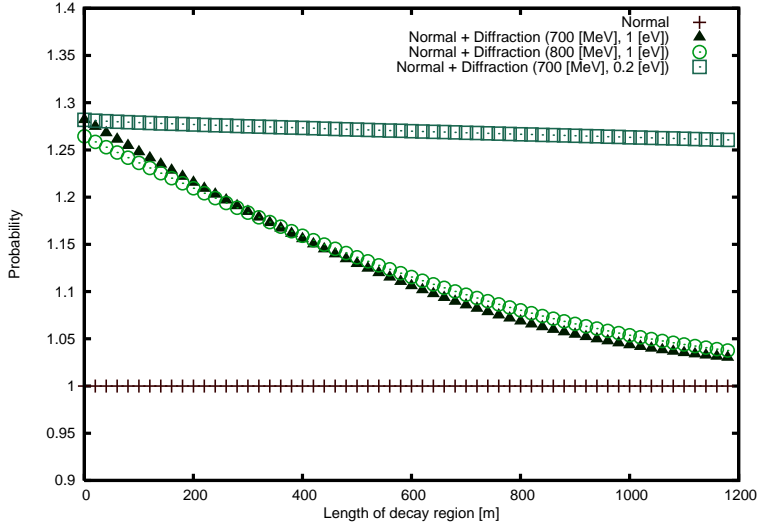}%
  \end{center}
  \caption{Total rate of the event that  the muon neutrino is detected at 
a finite
  distance $L$ for $\tau_{\pi}=\infty$.
 The constant (red line) shows the normal term,
 and the diffraction  term is given on top of the normal term.  The horizontal axis 
  represents the distance in meters, and  the normal term  is
  normalized to $1.0$.  The
  neutrino mass, pion energy, and neutrino energy are
  1.0 eV/$c^2$ or  0.2 eV/$c^2$, 4 GeV, and 700 (blue triangles) or ~800 (green
  circles) MeV, respectively. The excess varies with the distance  for
  $m_{\nu}=1.0$ eV/$c^2$ and is almost constant for
  $m_{\nu}=0.2$ eV/$c^2$.}
  \label{fig:2}
 \end{figure}%
Infinite number of
 states in $|\Psi(t) \rangle $ in Eq. $(\ref{pertubative-wave})$ of
 almost 
 identical  phases give   
the light-cone singularity to $\Delta_{\pi,l}(\delta x)$, and 
  the finite-size correction expressed in $\tilde
 g(\omega_\nu{T}) $.   
The present  quantum mechanical effect  remains at the 
macroscopic distance,  $2c\hbar E_{\nu}/(m_{\nu}^2c^4)$.

Next, we evaluate each term of Eq. $(\ref{probability-3})$. In  $G_0$, 
approximately, $p_{\pi} =p_l+p_{\nu}$, 
  and 
$2p_{\pi}\cdot p_{\nu}-m_{\pi}^2=m_l^2$. 
Integrating over the neutrino's angle, we find that this
  term is independent of  $\sigma_{\nu}$, which is consistent
 with  the condition for the stationary state \cite{Stodolsky}, and the 
  rate agrees with the value obtained by the ordinary method.
 In  
 $\tilde{g}(\omega_{\nu}{T})$,  $p_{\pi} \neq p_l+p_{\nu}$, 
and  the inner product, $p_\pi\cdot p_\nu$, is not expressed with  the masses of
 pion and charged lepton.
Instead, 
the convergence condition requires that this term is present in the
 kinematical region,
$|\vec{p}_{\nu}|(E_{\pi}-|\vec{p}_{\pi}|)\leq p_{\pi}\!\cdot\! p_{\nu}
\leq {\tilde{m}^2/2}$.
 
 It is 
impossible to experimentally distinguish both components, therefore,
 we add both terms. The total probability thus obtained 
is
 presented  in 
Fig. \ref{fig:2} for  neutrino masses 
 $m_{\nu}=1$ eV/$c^2$ and  $0.2$ eV/$c^2$, a pion
 energy $E_\pi = 4$ GeV of mean life-time $\tau_{\pi}=\infty$, and
 the  neutrino
 energies $E_\nu = 700$ and $800$ MeV. For the wave packet size of
 the neutrino,  we use the size
 of the nucleus having a mass number $A$, $\sigma_{\nu}=
 A^{\frac{2}{3}}/m_{\pi}^2$. 
For the ${}^{16}$O nucleus, $\sigma_{\nu}= 6.4/m_{\pi}^2$.
   From Fig. \ref{fig:2},
 we see   that  an excess  varies with the distance 
for  ${L}<1200$ m for $m_{\nu}=1$ eV/$c^2$ 
  and is almost constant for $0.2$ eV/$c^2$ and  that the
 maximal excess is approximately  $20$\% of the normal term at
 ${L}=0$. The 
slope at the origin ${L}=0$ is
 determined by $\omega_{\nu}$.
The diffraction term varies slowly  with both distance and energy.
For this situation, the typical length is  
${L}_0~[\text{m}] =2E_{\nu} \hbar c / (m_{\nu}^2c^4)= 400\times {E_{\nu}[\text{GeV}]/
 m_{\nu}^2[\text{eV}^2/c^4]}$.
 The neutrino's 
energy is measured experimentally with uncertainty $\Delta E_{\nu}$,
 which is of the 
order of $0.1 \times E_{\nu}$. This uncertainty is $100$ MeV for 
$1$ GeV neutrino energy and  the diffraction components of  both energies are
 almost equivalent  to those given in  Fig. \ref{fig:2}.
For the case of larger energy uncertainty, the computation  is easily done using
Eq. (\ref{probability-3}).

Using the asymptotic behavior of $\tilde g(\omega_{\nu} T)={2 \over \omega_{\nu}
T}$, we find an analytic expression of  the correction, although 
 the precise form of $\tilde g(\omega_{\nu} T)$ 
is used in comparing the theory with the experiments.  The 
rate,  is expressed with the 
$\Gamma_0$ of Eq. $(\ref{life-time})$ and 
this  correction as
\begin{align}
&\Gamma(T,\sigma_{\nu})=\Gamma_{0}+\Gamma^\text{diff}(T,\sigma_{\nu}) ,
\label{rate-general}\\
&\Gamma^\text{diff}(T,\sigma_{\nu})= \frac{G_F^2 f_{\pi}^2}{320 \pi^2}
 m_{\pi}^4\left(1-\frac{m_{l}^2}{m_{\pi}^2}\right)^4\left(1+\frac{4m_{l}^2}{
 m_{\pi}^2}\right){m_{\pi}^2 \sigma_{\nu} p_{\pi} \over T m_{\nu}^2}{1 \over E_{\pi}} .\nonumber
\end{align}
$\Gamma^\text{diff}(T,\sigma_{\nu})$ depends on $T$ and $\sigma_{\nu}$,
  and decreases with $T$ for a fixed  $\sigma_{\nu}$, and increases with
  $\sigma_{\nu}$ for  a fixed $T$. 
The  asymptotic value $ \lim_{ T
 \rightarrow \infty}\Gamma(T,\sigma_{\nu}) $ is independent of
 $\sigma_{\nu}$ and is computed with plane waves of
 $\sigma_{\nu}=\infty$, which agrees with those computed with
  $S[\infty]$ combined with $i\epsilon$ prescription.
It is noted that,
\begin{align}
&\lim_{\sigma_{\nu} \rightarrow \infty} \left\{ \lim_{ T
 \rightarrow \infty}\Gamma(T,\sigma_{\nu}) \right\} =\Gamma_{0} \label{limit1},\\
&\lim_{ T  \rightarrow \infty} \left\{\lim_{\sigma_{\nu}
 \rightarrow \infty}\Gamma( T,\sigma_{\nu})\right\} = \infty \label{limit2}.
\end{align}
The fact that   Equation $(\ref{limit2})$ diverges is consistent with the
behavior of the coefficient of $1/T$ term in Fermi's golden rule given
in Appendix.  $ c{T} \gg \sigma_\nu$ has been studied often and
  Equation $(\ref{limit1})$ is
applied.  $\sigma_\nu \gg c{T}$ has been less studied but 
there are  various places and   Equation $(\ref{limit2})$ is applied. Due to
diverging rate, intriguing phenomena may arise. 
 Thus   a careful 
consideration is necessary when  $(\sigma_\nu,T) \rightarrow
(\infty,\infty)$ is studied.
\subsubsection{Neutrino  spectrum vs charged lepton spectrum}
\begin{figure}[t]
\begin{center}
\hspace{-2cm}
 \includegraphics{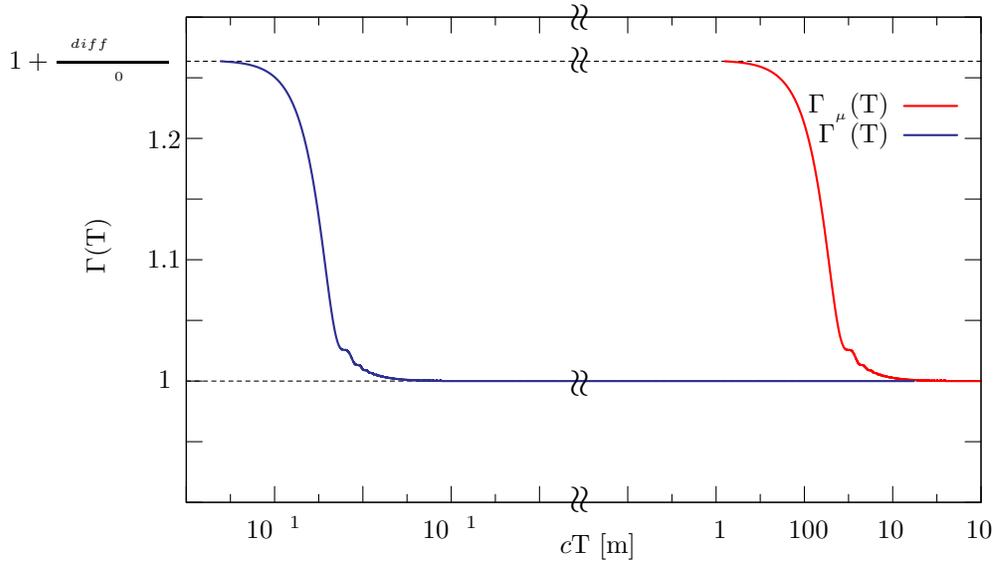}
\end{center}
\caption{The event rates  that   the  muon and muon neutrino are detected 
at a finite
  distance $L=cT$ for $\tau_{\pi}=\infty$.
 For the muon written with a blue line, the rate becomes   
the asymptotic value at $ 10^{-12}$ m, and 
for the neutrino written with a red line that becomes at $ 1000$ m.   The horizontal axis 
  represents the distance in meters, and  the asymptotic value $\Gamma_0$ is
  normalized to $1.0$.  The
  neutrino mass, pion energy, and neutrino energy are
$E_\pi=4$ GeV, $E_\nu=800$ MeV, $m_\nu = 1$ eV/$c^2$, $\sigma$:
 ${}^{16}O$. The detector's size is not considered.}
\label{fig:3}
\end{figure}%
In pion decays, the lepton and neutrino are produced in pair by the local 
weak interaction $H_\text{int}$ and the wave function $|\Psi(t) \rangle$  in 
Eq. $(\ref{pertubative-wave})$ expresses
the whole system, and the norm of 
parent wave function decreases with the average  life-time $\tau_{\pi}$ 
and that of the daughters increases with  
the same $\tau_{\pi}$. Experiments observe the events that the daughters are 
detected and the probability is computed with the amplitude that is defined 
according to this boundary condition, which becomes different at a finite $T$
 from    $\tau_{\pi}$ of Eq. $(\ref{pertubative-wave})$. Because the neutrino 
 has   almost the  constant phase and group velocity, its wave packet  
keeps coherence and    retains the interference 
pattern for long distance. When the neutrino   is  detected in this wave zone,
the event rate  is amplified of revealing  the  finite-size correction. The 
  leptons, on the 
other hand, are   massive and have the velocities that  vary with the 
momentum, and the wave packet does not 
show the strong effect.     
 \begin{figure}[t]
  \begin{center}
   \includegraphics[angle=0,scale=1.0]{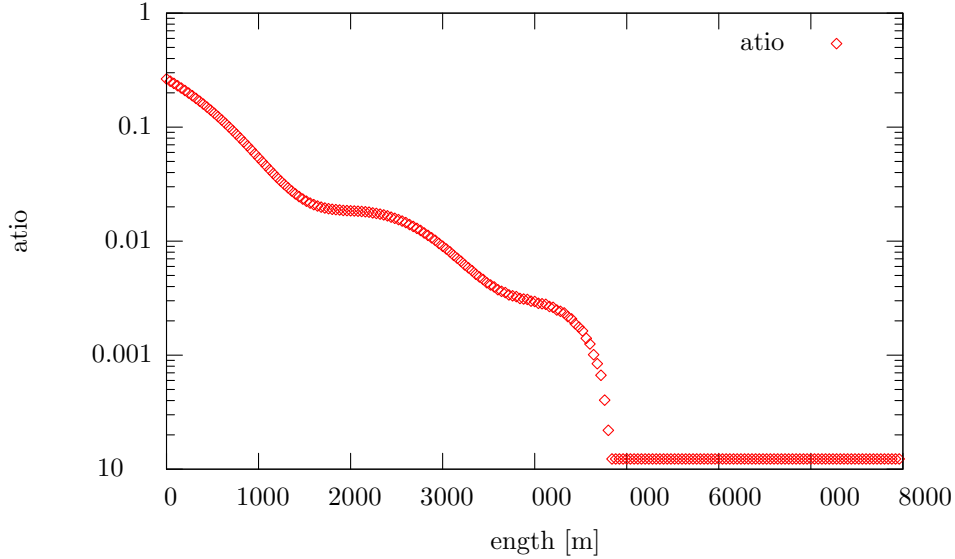}%
  \end{center}
  \caption{The ratio of the event rate Eq. $(\ref{nue/numu})$ that  the
  electron neutrino is detected over that  the muon neutrino is
  detected. The value varies from $0.2$ at $\text{L}=0$ to 
$10^{-4}$ at $\text{L}=\infty$ slowly.    The horizontal axis 
  represents the distance in meters.   The
  neutrino mass, pion energy, and neutrino energy are
  1.0 eV/$c^2$ , 4 GeV, and 800 MeV, respectively. 
The detector's size is not considered.}
  \label{fig:nue over numu}
 \end{figure}%
In fact,   $\Gamma^{diff}(T,\sigma_{i});i=\nu,l$  in  Eq. $(\ref{rate-general})$
 is inversely proportional to 
the  mass-squared and is large for the neutrino $i=\nu$ and is small for 
the charged leptons $i=l$
 if
$\sigma_{l} \approx \sigma_{\nu}$. Thus the rates at $T=0~\text {and}~ \infty$ 
, for $\sigma_{\nu}=\sigma_l=\sigma$, satisfy
\begin{align}
&\Gamma(T=\infty,\sigma_{\nu}=\sigma;neutrino)=\Gamma(
 T=\infty,\sigma_{l}=\sigma;lepton),  \\
&\Gamma(T=0,\sigma_{\nu}=\sigma;neutrino)=\Gamma(
 T=0,\sigma_{l}=\sigma;lepton),
\end{align}
but at finite  $T$, they are different.  
Figure \ref{fig:3} shows them. The finite-size correction appears  even at a macroscopic $T$ for the neutrinos, and appears only
 at a microscopic $T$ for the charged leptons.  Even though they are produced 
in pair, they propagate differently and  are detected with the  different
rates due to the finite-size corrections. 
The  ratio varies from $10^{-4}$ at $ T=\infty$ to $0.2$ at  
$T=0$. The enhancement of electron mode at finite $T$ is huge.
\subsubsection{Suppression of electron  mode}
The ratio of the event that  the electron is detected  over 
the event  that  the muon is detected  at $T=\infty$  is
\begin{eqnarray}    
\label{helicity-suppression}
\Gamma_{0}(electron)/\Gamma_{0}(muon)={m_e^2 \over m_{\mu}^2} \left(1-{m_{\mu}^2 \over m_{\pi}^2}\right)^{-2}
=1.28 \times 10^{-4},
\end{eqnarray}
which is consistent with the experimental value $1.23 \times 10^{-4}$. 
The  suppression of the electron mode played the important role to 
prove the form of interaction to 
 $V-A$ type.  The branching ratio vanishes for massless charged lepton 
because they have opposite
helicities and decouple from scalar
or pseudo-scalar particle. Thus the conservation law of kinetic energy 
and angular momentum, which hold at $T=\infty$, suppresses  the 
electron mode. Now  the finite-size 
correction comes from the final states of the non-conserving  
kinetic energy, hence does not follow  the helicity suppression. The 
probabilities of the events that the charged leptons  are detected at 
$c T=10^{-10}$ m agree with  the asymptotic values of
Eq. $(\ref{helicity-suppression})$. Whereas, the  ratio 
of the probabilities of the events that the electron neutrino is
detected over that the muon neutrino is detected  at a $T$,
\begin{eqnarray}    
R_e(T)=\Gamma(T,\nu_e )/\Gamma(T,\nu_{\mu})
\label{nue/numu}
\end{eqnarray}
becomes very different from Eq. $(\ref{helicity-suppression})$ due to the large
finite-size corrections, $\Gamma^\text{diff}( T,\sigma)$ in
Eq. $(\ref{rate-general})$. The ratio  of the total number of events,
which is slightly different from the experimental value  due to the 
finite size of the detector, at finite $T$ for  1.0 eV/$c^2$ , 4
GeV, and 800 MeV for the  
  neutrino mass, pion energy, and neutrino energy is shown in
  Fig. \ref{fig:nue over numu}. In Fig.  \ref{fig:nue over numu}, the ratio
of   probabilities integrated over the whole angles are plotted, and is
considered the maximum value. 
   
The  ratio varies from $10^{-4}$ at $ T=\infty$ to $0.2$ at  
$T=0$. The enhancement of electron mode at finite $T$ is huge.
In real experiments, the detectors of
finite sizes are used. The values then become smaller, and 
become consistent with  existing values within experimental
uncertainties. Later we compare theoretical values with experimental data,
 then the size and geometry of detector are included. 
 
\subsection{Intermediate-energy  pion } 

So far we have studied the high-energy pion  whose 
life-time is ignorable, and one flavor neutrino.   Hereafter the
pion's life-time and three flavor  are included.
The pion's  life-time  gives the damping factor $e^{-\frac{t_1+t_2}{\tau_{\pi}}}$ to the
left-hand side of Eq. $(\ref{probability1})$. The new
 universal   
function $\tilde g(\omega_\nu, {T};\tau_\pi)$
\begin{align}
&i  \int_0^{{T}} dt_1 dt_2  \frac{\epsilon(\delta t)}{|\delta
 t|}e^{i {\omega_{\nu}}\delta t -{t_1+t_2 \over \tau_{\pi}}}   
=  \tilde g(\omega_{\nu},{T};\tau_{\pi})-\tilde g_0;\  T>{\sqrt \sigma_{\nu} \over c},
\label{probability2} 
\end{align}
where $\tilde g(\omega_{\nu},\infty,\tau_{\pi})=0$, replaces 
$T \tilde g(\omega_\nu {T})$, and $(1-e^{-{T}/\tau_{\pi}})G_0$ replaces
$T G_0$.

The
integrand of Eq. $(\ref{probability2})$  proportional to  
\begin{align}
\exp\left[ (i \omega_{\nu}  -{1/\tau_{\pi}})t_1 \right]\times \exp\left[
(-i\omega_{\nu} -{1/\tau_{\pi}})t_2\right],
\end{align}
 shows that a motion is   equivalent to a
damped oscillator of the angular velocity $\omega_{\nu}$ and the decay
rate ${1 \over \tau_{\pi}}$. 
Now    for $E_{\nu}=1$ GeV their values are, 
\begin{align}
\hbar \omega_{\nu} &=\begin{cases}
		      10^{-9}\ \text{eV},
\ \text{for} ~m_{\nu}=1 \ \text{eV}/c^2, \\
 10^{-11}\ \text{eV},\ \text{for} \ m_{\nu}=0.1\ \text{eV}/c^2,
		     \end{cases}\label{diffraction-length}\\
{\hbar \over \tau_{\pi}}&=\begin{cases}
			   3\times 10^{-8}\ \text{eV},\ \text{at rest}, \\
			   1.5 \times 10^{-10}\ \text{eV},\ E_{\pi}=50 m_{\pi}c^2.
			  \end{cases}\label{life-length} 
\end{align}
The motion  is sensitive to $\omega_{\nu}$ in the
region of  $ \omega_{\nu} \approx {1 \over \tau_{\pi}}$. Now $\omega_{\nu}$ is proportional to the square of the absolute value of neutrino mass, thus $P$   
is useful to probe it around  $m_{\nu}=0.1$ {eV}/$c^2$.   Mass-squared differences $\delta m^2_\nu$ are   
extremely  small   \cite {pdg,Tritium,WMAP-neutrino}, and
their 
central value is currently unknown. If that is in the above range,  the 
probability  in the region $T \leq \tau_{\pi}$  can be used to measure 
 the absolute neutrino mass.  
 \begin{figure}[t]%
  \begin{center}
   \includegraphics[angle=-90,scale=.25]{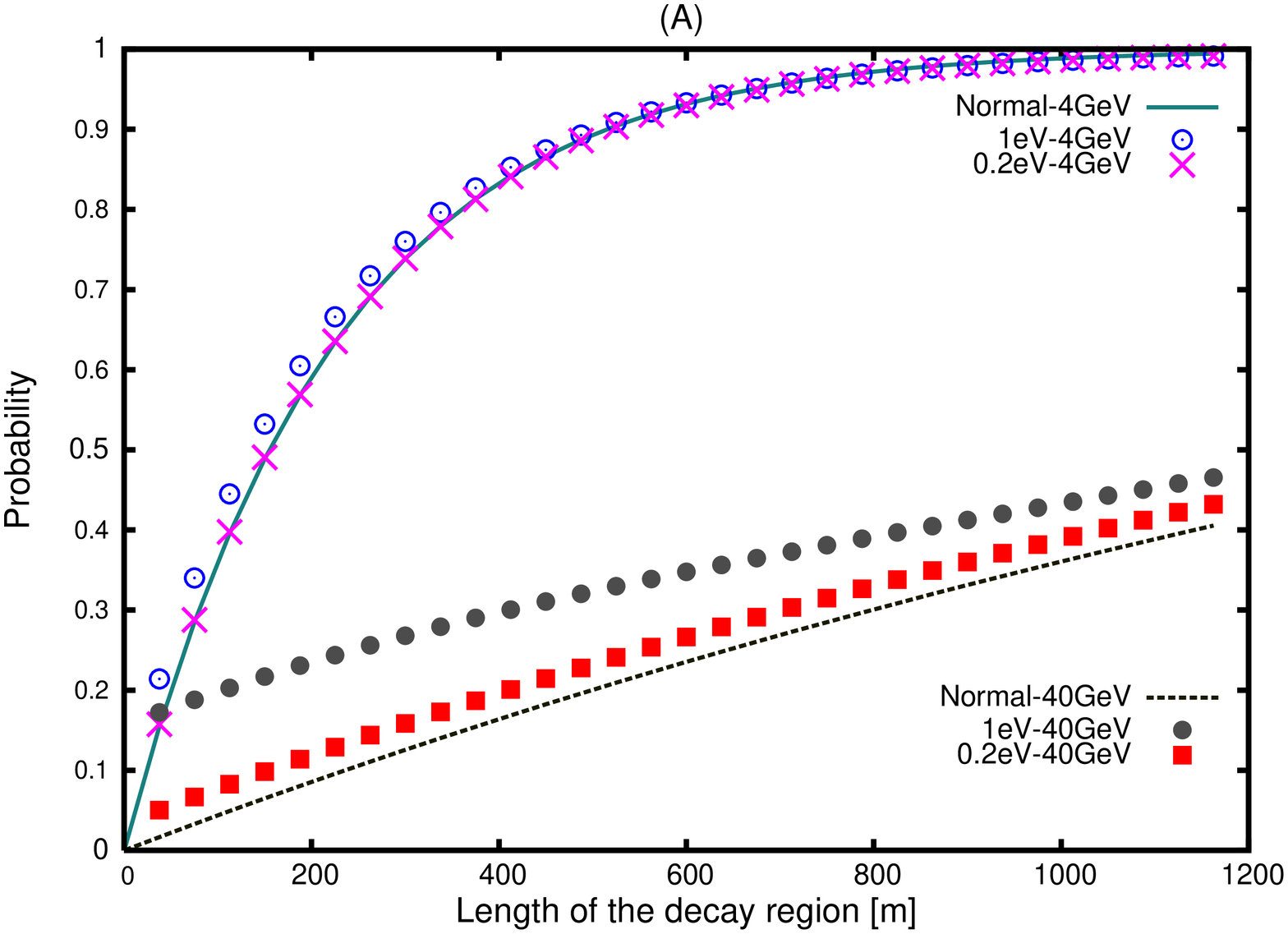}
   \includegraphics[angle=-90,scale=.24]{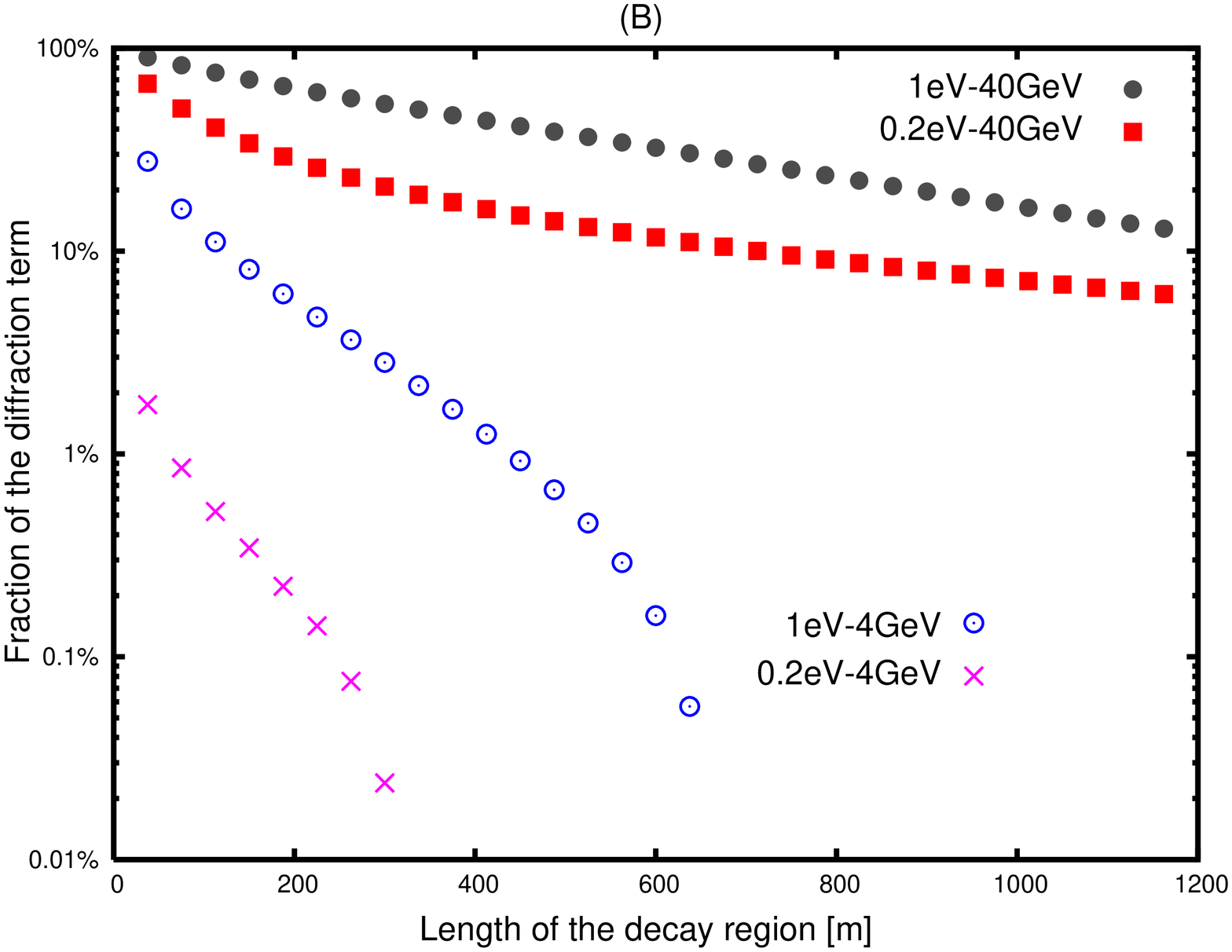}%
  \end{center}
  \caption{Probability  of the event that the neutrino is  detected  at   
distance $L$ for $E_{\pi}=4$ and $40$ GeV.   In (A),  solid  green ($4$ GeV) 
and  dotted black line  ($40$ GeV) represent the normal  and
  diffraction 
terms, respectively, for  for 
$m_{\nu}$ of $0.2$ and $1.0$ eV/$c^2$. The diffraction terms are written   on 
top of  the normal
  terms. Values   are normalized 
to one  at ${L}=\infty$.
 In (B),   fractions of 
  diffraction terms that    vary with
   pion's energy and  neutrino mass are shown. The horizontal axis 
  represents  distance in~[m]. 
   Neutrino energy is 700 MeV.
The detector's size is not considered.}
  \label{fig:5}
 \end{figure}%

The 
probabilities  of the events for  several parameters are given in
Fig. \ref{fig:5}.   The finite-size correction  becomes smaller 
 for  lower energy and  the life-time effect becomes significant  for smaller
 neutrino mass.  Thus the  fraction 
 becomes sensitive to the neutrino mass and negligibly small in 
low energy pion in ${L} > 200$ m, if the mass is less 
than $0.05$ eV/$c^2$. 
For three neutrinos of masses $m_{{\nu}_i};i=1-3$ and a mixing matrix
 $U_{i,\alpha};\alpha=e,\mu,\tau$, the Lagrangian Eq. $(\ref{lagrangian})$ is modified to  that of three neutrino $\nu_{i}(x);i=1-3$ of the masses
 $m_{{\nu}_i}$ and the weak current 
\begin{eqnarray}
J_{V-A}(x)= {J_{V-A}}^{i}(x) U_{i,\alpha}.
\end{eqnarray}
The amplitude for the neutrino of flavor 
$\alpha$ and the charged lepton $l$ is given by 
\begin{eqnarray} 
\mathcal{M}=\sum_i U_{l,i}\mathcal{M}(l,i)U_{i,\alpha}^{\dagger},
\end{eqnarray} 
and  the probability  is
\begin{eqnarray} 
P&=&\sum_{l=e,\mu} \int d{\vec X}_{\nu_{\alpha}}{d{\vec p}_{\nu_{\alpha}} \over (2\pi)^3}
{d{\vec p}_l \over (2\pi)^3} |\mathcal{M}|^2 \nonumber 
\\
 &=&\sum_{l=e,\mu}\int d{\vec X}_{\nu_{\alpha}}{d{\vec p}_{\nu_{\alpha}} \over (2\pi)^3}
{d{\vec p}_l \over (2\pi)^3}|U_{l,i}\mathcal{M}(l,i)U_{i,\alpha}^{\dagger}|^2, \label{three-f-probability}
\end{eqnarray} 
where $\mathcal{M}(l,i)$ is the amplitude for the mass eigenstate $i$. 
$P$ of Eq. $(\ref{three-f-probability})$ includes  the diffraction 
and normal terms. The diffraction term exists  in the kinematical region of 
${\vec p}_{\nu_{\alpha}}$ satisfying the condition    Eq. $(\ref{convergence})$,
 which  depends on the charged lepton's mass, and is independent of 
the  species  in its form. Thus  
the diffraction term in 
\begin{eqnarray}
\int \frac{d{\vec p}_l}{(2\pi)^3}\mathcal{M}(l,i)\mathcal{M}^{*}(l,j)
\label{l-i-diffraction}
\end{eqnarray}
does not depend on $l$. The phase   of the 
integrand  in Eq. $(\ref{probability2})$ is replaced with  the  phase,  
${\omega_{\nu_i}+\omega_{\nu_j} \over 2}(t_1-t_2)+\frac{\omega_{\nu_i}-\omega_{\nu_j}}{2}(t_1+t_2)$. The latter  is much smaller than 
${1 \over \tau_{\pi}}(t_1+t_2)$ in  magnitude and  can be ignored in the short distance region. 
Hence   the diffraction in  Eq. $(\ref{l-i-diffraction})$ is proportional to 
$\tilde g({\omega_{\nu_i}+\omega_{\nu_j} \over 2},T;\tau_{\pi})$. The normal 
term satisfies the conservation law of kinetic energy and momentum, and  
$m_{\pi}^2-2p_{\pi}\!\cdot\!p_{\nu_{\alpha}}=m_l^2$,  hence the integrand in 
electron mode is negligible. Thus $l=\mu$ contributes. This  
depends on
$i$  through  $D_i({\vec p}_{\nu},
T)=e^{i{(E_{\nu_i}({\vec p}_{\nu})T-p_{\nu}c T)}}$  and 
$U_{i,\alpha}$, and agrees with the standard oscillation formula.

In the kinematical region $V_1:2p_{\pi}\!\cdot\! p_{\nu} \leq m_{\pi}^2-m_{\mu}^2$, 
the convergence condition  Eq. $(\ref{convergence})$ for $l=e, \mu$ is satisfied, and 
we have  the probability of the event that   the neutrino of flavor
$\alpha$ is detected, 
\begin{eqnarray}
P&=&\frac{N_4}{T} \int_{V_1} {d\vec{p}_{\nu_{\alpha}} \over (2\pi)^3} 
 \frac{p_{\pi}\! \cdot\! p_{\nu_{\alpha}}(m_{\pi}^2-2p_{\pi}\! \cdot\!
 p_{\nu_{\alpha}}) }{E_{\nu_{\alpha}}}\sum_i
\nonumber\\
& & \times \left[ \sum_{j,l} U_{ij}^{\alpha} (U_{ji}^{l})^{*} \tilde g(\omega_{i,j},{T};\tau_\pi) 
 +|U_{l,i} D_i({\vec p}_{\nu_{\alpha}},T)
 U^{\dagger}_{\alpha,i}|^2 G_0(1-e^{-T/\tau_\pi}) \right],\label{three-flavor}
\\
& &U_{ij}^{\alpha}= U_{i,\alpha}^{\dagger} U_{j,\alpha},\ \omega_{ij}={\omega_{\nu_i}+\omega_{\nu_j} \over 2},\nonumber
\end{eqnarray}
where  
$E_{\nu_{\alpha}}=p_{\nu_{\alpha}}$ can be  used except in 
$D_i({\vec p}\,)D_j^*({\vec p}\,)=
e^{i\frac{m_{\nu_i}^2-m_{\nu_j}^2}{2p}{T}} $.

In the kinematical region $V_2:m_{\pi}^2-m_{\mu}^2 \leq 2p_{\pi}\cdot p_{\nu}
 \leq m_{\pi}^2-m_{e}^2$, the convergence condition Eq. $(\ref{convergence})$ 
for $l=e$ is
 satisfied, and  the diffraction gives  a contribution. The normal term
 is negligible.
 Thus we have
\begin{eqnarray}
P=\frac{N_4}{T}  \int_{V_2} {d\vec{p}_{\nu_{\alpha}} \over (2\pi)^3} 
 \frac{p_{\pi}\! \cdot\! p_{\nu_{\alpha}}(m_{\pi}^2-2p_{\pi}\! \cdot\!
 p_{\nu_{\alpha}}) }{E_{\nu_{\alpha}}} \sum_{i,j} U_{ij}^{\alpha} (U_{ji}^{e})^{*} 
\tilde{g}(\omega_{ij},{T};\tau_\pi) ,
\label{three-flavor2}
\end{eqnarray}

Equations. $(\ref{three-flavor})$ and $(\ref{three-flavor2})$ are the formula
 for the probability of the event
of the neutrino $\alpha$ at the distance $c{T}$. The second term in the right-hand side of Eq. $(\ref{three-flavor})$ is  the standard
 flavor oscillation term that depends on the mass-squared difference. The rests 
are  
 the diffraction terms that depend on the square of average mass, 
$\omega_{ij}={ m_{\nu_i}^2+ m_{\nu_j}^2 \over 2}$. It is convenient to 
 define the average squared-mass ${\bar m_{\nu}}^2$ from 
\begin{eqnarray}
 \tilde{g}(\bar{\omega}_{\nu},{T};\tau_{\pi})= \sum_{i,j,l} U_{ij}^{\alpha} (U_{ji}^{l})^{*}   \tilde{g}
  (\omega_{\nu_i},{T};\tau_\pi),\
  \bar{\omega}_{\nu}={{\bar{m}_{\nu}}^2 \over 2E_\nu},
\label{average-mass}
\end{eqnarray}
where  ${\bar m_{\nu}}$ agrees with the central mass $m_0$ if  
$m_0^2 \gg \delta m^2_\nu$.
At ${L} \leq  c\tau_\pi $, the first term gives a large contribution 
for $\alpha= e$ , but a small correction for $\alpha= \mu$. At ${L} \gg c\tau_\pi $, the
diffraction term disappears and  the expression agrees with the standard
  flavor oscillation formula. 
\begin{figure}[t]
\includegraphics[scale=.3,angle=-90]{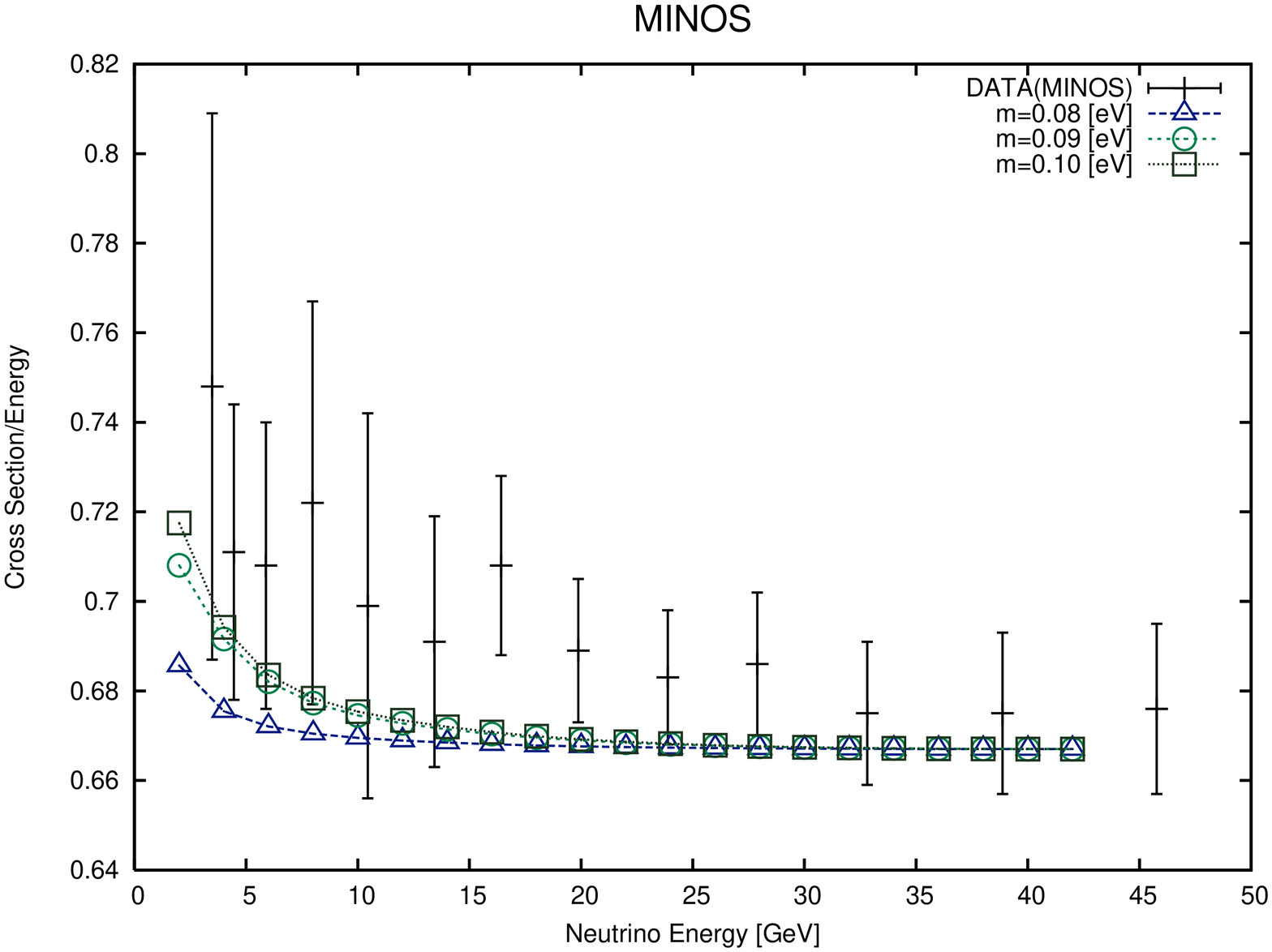}
\includegraphics[scale=.3,angle=-90]{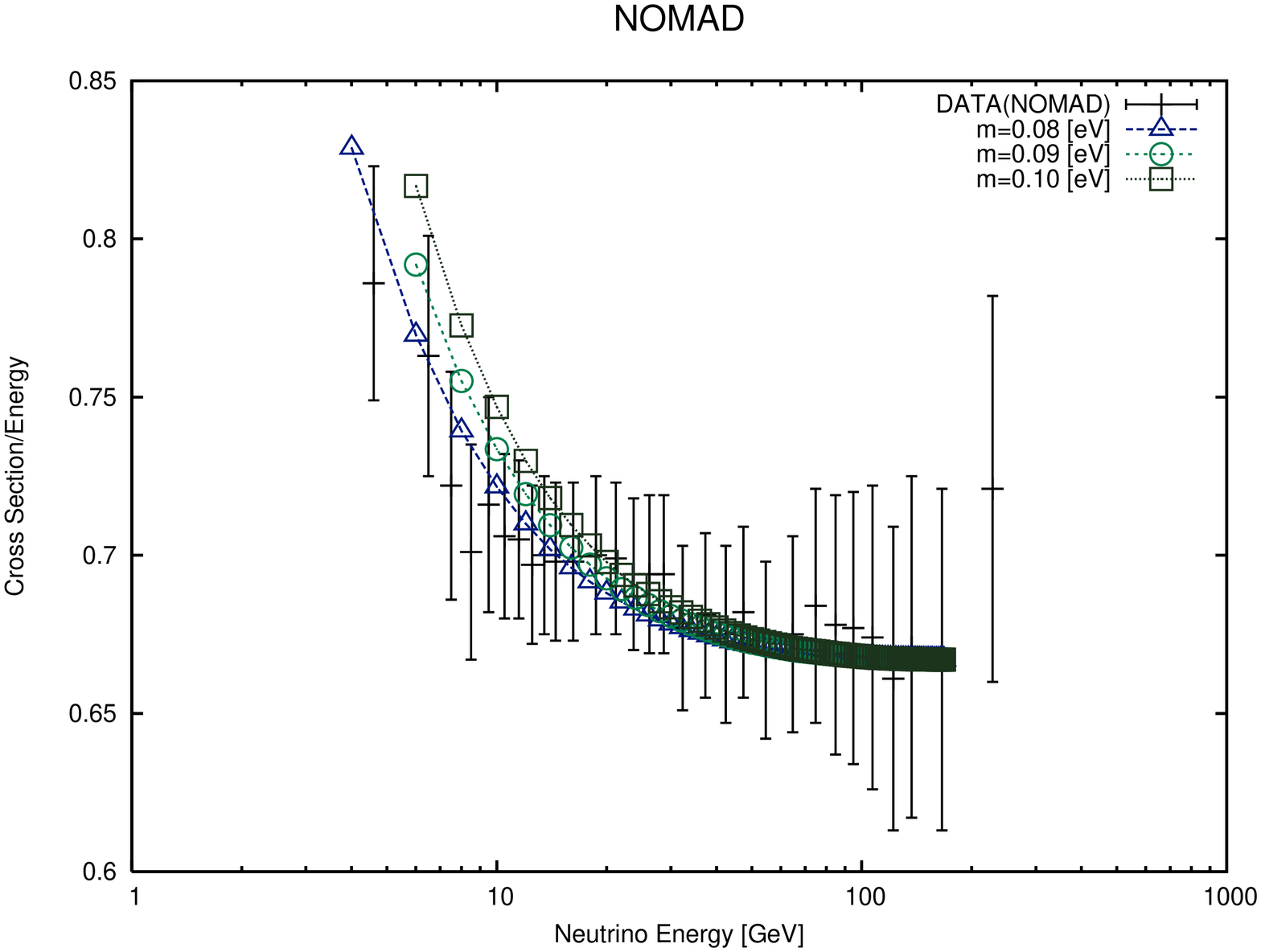}
\caption{$\nu N$ total cross section from MINOS  and NOMAD collaborations  are compared with theory.
Theory uses $\sigma_{\nu}= 14.6/m_{\pi}^2 $ of \ ${}^{56}$Fe, and geometry of   MINOS 
and $\sigma_{\nu} =5.3/m_{\pi}^2$ of  ${}^{12}$C, and geometry of  NOMAD. 
 The horizontal axis gives neutrino energy in GeV and the vertical
 axis gives the ratio of cross section to energy. The neutrino mass, 
$m_{\nu}=0.08,\ 0.09,\ 0.10$ eV/$c^2$, is used for theoretical calculation.}
\label{fig:6}
\end{figure}

Now we compare experimental data  with the theoretical values  Eq. $(\ref{three-flavor})$. In Fig. \ref{fig:6}, the 
total cross sections of 
high energy neutrino-nucleon scattering of the NOMAD
\cite{excess-total-detectorNOMAD} and MINOS \cite{excess-total-detectorMino} 
collaborations are presented. The geometry 
of each experiment is taken into account in the theoretical
calculation and  ${\sigma}_{\nu}= 5.3/{m_{\pi}^2}$ or $14.6/{m_{\pi}^2}$  
of  ${}^{12}$C and 
${}^{56}$Fe  
are used. 
The  theoretical values depend slightly on the neutrino mass 
and energy.  In these high energy regions, the life-time 
of the parent does not give a significant effect and the corrections  
 remain in the distances of these experiments.  They give  
slight energy dependences to the total cross sections 
and agree with experiments. Other experiments except those  discussed
next are not in-consistent with the correction terms within experimental 
uncertainties. 
\begin{figure}[t]
\begin{center}
\includegraphics[scale=.4,angle=-90]{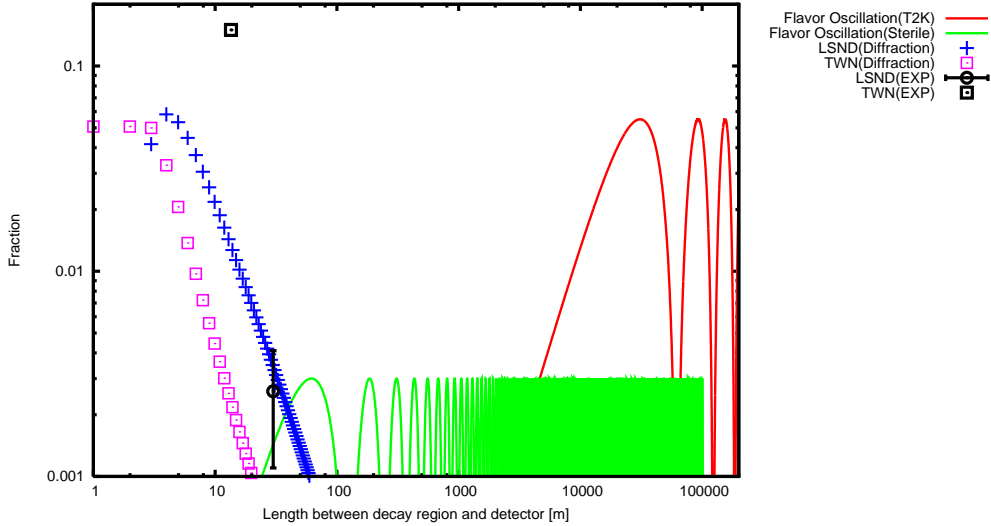}
\end{center}
\caption{Theoretical  fractions of the electron mode 
  from the diffractions for LSND and TWN geometries are compared with
   LSND and TWN
  experiments.  TWN(black box) and  LSND(black circle) are  the
  experimental values and  magenta  line of boxes is 
the value from diffraction  for TWN of  the
  parameters $m_\nu=0.2$ eV/$c^2$,
	    $E_\nu=300$ MeV, and  $p_\pi=1.4$ GeV/$c$.
	    Blue cross line  is the value for LSND with  
$m_\nu=0.1$ eV/$c^2$, $E_\nu=60$ MeV,
	    and $p_\pi=400\,\text{MeV}/c$. Red line shows the value
 for flavor oscillation (T2K)  for 
	    $\sin^2\theta_{13}=0.11$, $\delta m^2_{23} =
	    2.4\times10^{-3}\ \text{eV}^2/c^4$, and $E_\nu = 60$ MeV.
 Green solid shows the value for the sterile neutrino  
for $\sin^2\theta=0.004$, 
$\delta
	    m^2 = 1.2 \ \text{eV}^2/c^4$, and $E_\nu =
	    60$ MeV.
}
\label{fig:7}
\end{figure}
\subsection{Neutrino mass from LSND}  

Next an anomaly on the fraction  of electron mode over muon mode in the 
pion decay   is compared. At ${L}/c \gg \tau_\pi$, the flavor diagonal  
term has  the fraction  of $10^{-4}$  \cite{Sasaki,Jack,Ruderman,Anderson} and 
the value is determined by the flavor mixing  term in Eq. $(\ref{three-flavor})$.
 Neutrino parameters   determined   in Ref. \cite{pdg}  as
\begin{align}
&|\delta m^2_{23}| =
	    2.35^{+0.12}_{-0.09}\times10^{-3},~\delta m^2_{21} =
	    7.58^{+0.22}_{-0.26}\times10^{-5}\ \text{eV}^2/c^4, 
\ \sin^2\theta_{13}=0.096\pm 0.013,\label{flavour-parameter}
\end{align}
are used.
  In the short distance region, the electron mode due to   the diffraction   
has a significant
 fraction.  
Its magnitude    is about
0.1 to $10^{-4}$   depending the neutrino mass  and the distance  
from Figs. \ref{fig:nue over numu} and \ref{fig:5}.   Thus,
the excess   becomes  enormous.  Figure \ref{fig:7} shows
the 
fraction  that includes geometrical configurations  of
detectors. The value is about $0.05$ at a 
distance of a few meters  and  $0.01$ at a distance of 10 to 100
meters. That becomes negligible in a distance above 1000 meters. Thus  the 
diffraction gives the dominant contribution   in the 
short distance region and  the flavor oscillation gives the dominant 
contribution in the long distance region.

\begin{figure}[t]
\begin{center}
\includegraphics[scale=.45,angle=-90]{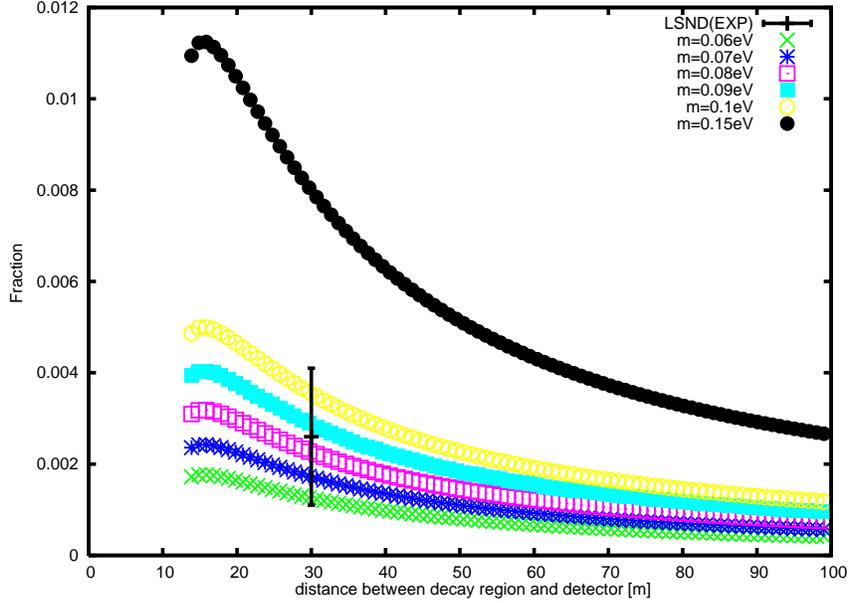}
\end{center}
\caption{Relative fraction of the electron mode computed theoretically
  for LSND geometry is compared with the experiment. The neutrino mass is from $0.06$ {eV}/$c^2$ to $0.15$ {eV}/$c^2$, and $\sigma_{\nu}=5.3/{m_{\pi}^2}$ from ${}^{12}$C.
The energies are  
	    $E_\nu=60$ MeV, and  $p_\pi=400$ MeV/$c$. 
    LSND (EXP)  is consistent with 
  the mass  $m_\nu = 0.082 ~ \pm 0.020$ eV/$c^2$.}
	   
\label{fig:8}
\end{figure}

\begin{figure}[t]
\includegraphics[scale=.3,angle=-90]{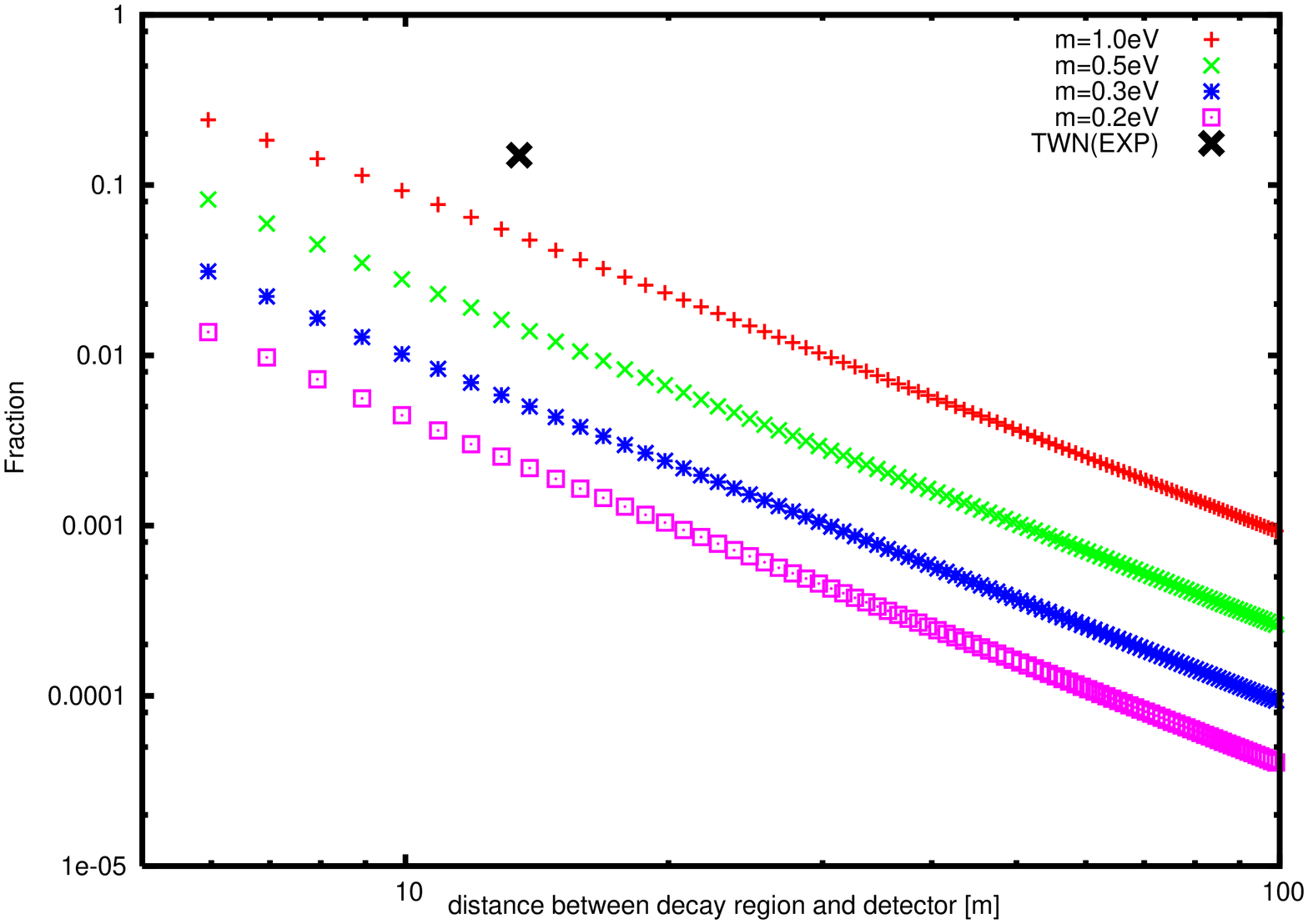}
\includegraphics[scale=.3,angle=-90]{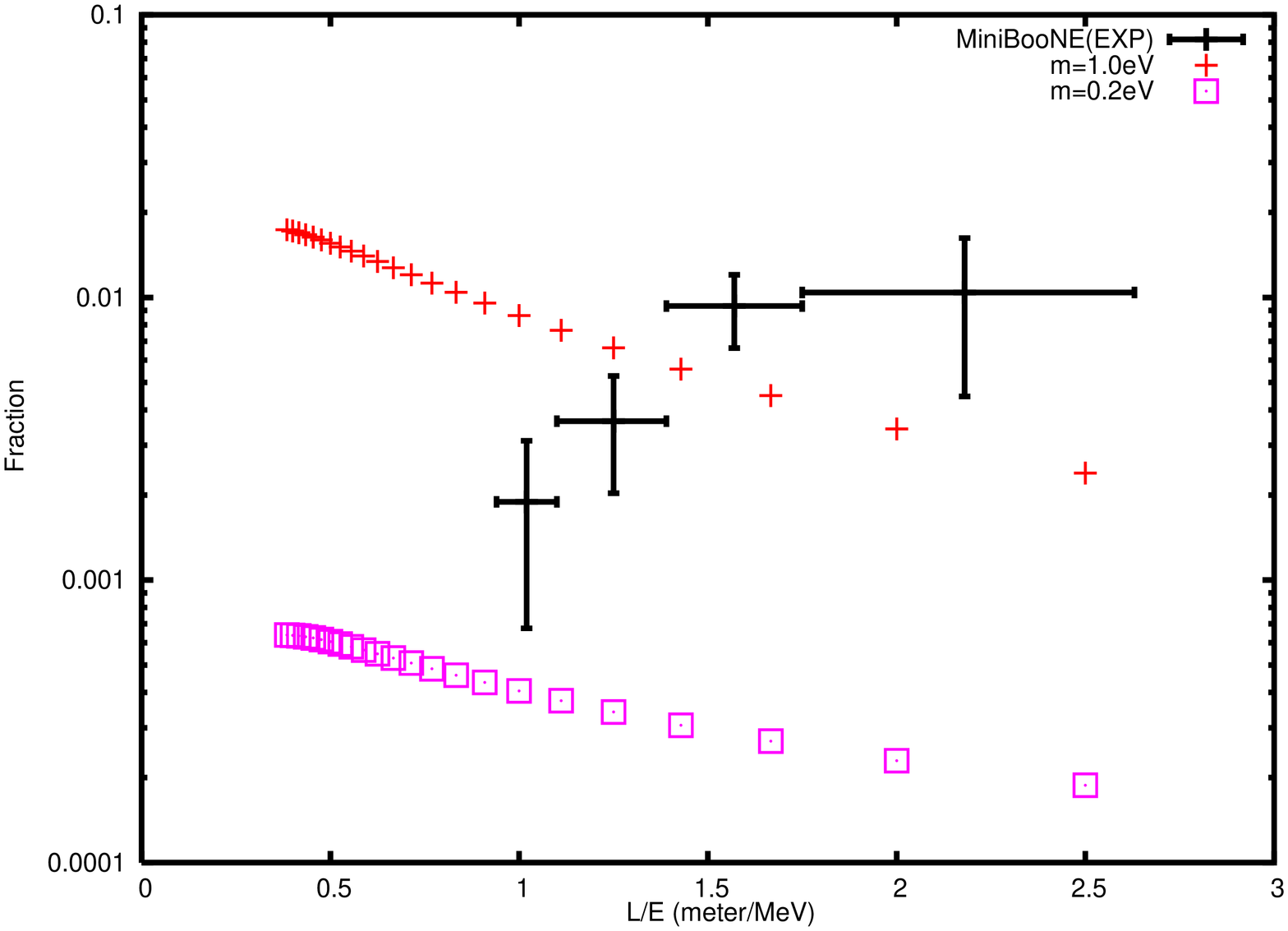}
\caption{Relative fraction of the electron mode computed theoretically
  for TWN and MiniBooNE geometries are  compared with
   experiments.  The neutrino mass is from $0.2$ {eV}/$c^2$ to $1.0$
  {eV}/$c^2$, and $\sigma_{\nu}=8.8/{m_{\pi}^2}$ or
  $5.3/{m_{\pi}^2}$ from ${}^{26}$Al or${}^{12}$C.
  The error of TWN  value is unknown, and the background estimation
  of MiniBooNE will be altered with the diffraction of the muon decay,
 hence the
  data  are  consistent with 
 the theoretical values  calculated with  $m_\nu \approx 0.082 ~ \pm
  0.020$ {eV}/$c^2$,
	    $E_\nu=300$ MeV,  $p_\pi=1.4$ GeV/$c$ for
  TWN and $E_\nu=200-800$ {MeV},  $p_\pi=3$ {GeV}/$c$ for MiniBooNE.
}
\label{fig:9}
\end{figure}

There have been several experiments in short distance regions. The first high-energy 
neutrino experiment, TWN  \cite{excess-two-neutrino},  and the 
first liquid-scintillator experiment, LSND  \cite{LSND}, detected the neutrinos in this region and modern 
experiments followed. TWN and LSND are compared with the theories of the flavor oscillations
 and the neutrino diffraction in Fig. \ref{fig:7}. 
The experiments of TWN and LSND are not consistent with the flavor oscillations. For the 
LSND experiment, the distance
is about 30 meters and the  fraction of the electron neutrino 
and those  of the flavor oscillation and the diffraction are plotted  in 
Fig. \ref{fig:7}. The  oscillation  probability with  the 
mass-squared difference of Eq. $(\ref{flavour-parameter})$ (red line) vanishes, and that with 
a much   larger value $\delta m^2=1.2\ \text{eV}^2/c^4$ (green line)
agrees with the experiment. Obviously this value is much larger than the 
values of global fit for three neutrinos.  Now  the diffraction 
component (blue line) with $\sigma_{\nu}=5.3/{m_{\pi}^2}$ from
${}^{12}$C used in the target  agrees with the experiment with 
the absolute neutrino mass
around $0.08$ eV/{$c^2$}.   Thus, we 
consider  that the LSND event is  a signal of neutrino diffraction. 
In Fig. \ref{fig:8}, we compare the experiment with the
theoretical values with the above wave packet size.  
They agree with the neutrino mass $\bar m_{\nu}=0.082\pm 0.020$ eV/$c^2$.
 MiniBooNE \cite{excess-near-detectorMini}
tested LSND later at a different distance $L=490$ m and higher energy 
and did not see 
the signal expected from  the flavor oscillation hypothesis of LSND. At 
this distance, 
the diffraction component with $\sigma_{\nu}=5.3/{m_{\pi}^2}$  from
${}^{12}$C for MiniBooNE becomes much smaller than that of LSND. Recent 
new data from  MiniBooNE \cite{excess-near-detectorMini-n}
claimed the anomaly in $L/E$ plot, and are compared  in 
Fig. \ref{fig:9}. The value  is slightly larger than the diffraction 
of $\bar m_{\nu}=0.082\pm
0.020$ eV/$c^2$.
However, the diffraction of neutrino from muon decay was not included in  
the background estimation. The present value, hence, is consistent with 
the diffraction of $m_{\nu}=0.082\pm 0.020$ eV/$c^2$. Thus    MiniBooNE's new 
and old data  are  consistent with the diffraction.
 The 
large signal  for LSND and small signal for MiniBooNE  are 
 consistent with the neutrino diffraction.

Excess of shower events in TWN \cite{excess-two-neutrino}, which may be caused by the 
electron neutrino, are compared  in 
Figs. \ref{fig:7} \text {and} \ref{fig:9}. The fraction  of 
shower events of  TWN is slightly  larger than those of theory.  
However,   shower events in TWN might include  the electron neutrino from the 
muon decay, which is furthermore amplified by  the neutrino diffraction, 
\cite{ishikawa-nozaki-tobita}. Hence   the 
experimental values 
are consistent   with the theory  using  the mass 
obtained from LSND  from  Fig. $\ref{fig:9}$.  

MiniBooNE for the decay
at rest  and reactor experiments  in a short 
distance also observed
 anomalies, which  may be attributed to the neutrino diffraction. Nevertheless  
the precise quantitative analysis are not straightforward,
 since  matter effects are more involved in these processes, and they  will
 be studied in a separate publication.  
Further experiments at shorter distance may be able to 
confirm the diffraction of the electron neutrino.  
  
From Eq. $(\ref{flavour-parameter})$, $\sqrt {\delta
m_{23}^2}$ is slightly smaller than the average value, $0.082$ eV/{$c^2$}, 
and is larger than the error  $0.020$ eV/{$c^2$},
and $\sqrt {\delta
m_{21}^2}$ is much smaller than both.
Hence the pattern of masses are either the normal or inverted  hierarchies.

Now we substitute the mixing matrix $U$  determined 
from the global fits \cite{pdg},
to Eq. $(\ref{average-mass})$ and have  the light
and  heavy masses,
\begin{align}
&m_{\nu_L}= 0.085\pm0.022\ \text{eV}/{c^2},\ m_{\nu_H}= 0.098 \pm0.022 \ \text{eV}/{c^2};\ \text{I},\label{masses1}\\
&m_{\nu_L}=0.070\pm0.026\ \text{eV}/{c^2},\ m_{\nu_H}=0.083\pm0.026\ \text{eV}/{c^2};\ \text{II},\label{masses2}
\end{align}
and the sum of masses   
\begin{align}
&   \sum_i m_{\nu_i}=0.268\pm0.078\ \text{eV}/{c^2};\ \text{I},\label{sum-masses1}\\
&  \sum_i m_{\nu_i}=0.236\pm0.066\ \text{eV}/{c^2};\ \text{II}, \label{sum-masses2}
\end{align} 
for the normal   (I) or inverted   (II) mass hierarchies. 

Thus the  probability of the detection process of the pion decay at 
the distance ${L}=cT$ 
is computed   by $S[{T}]$ as  Eqs. $(\ref{probability-3})$ 
or  $(\ref{three-f-probability})$, and 
deviates from  that at ${L}=\infty $ given by $S[\infty]$.
The probability  has    the large finite-size 
correction.  If the charged lepton is observed simultaneously, the
 detection rate of the lepton  has also the large finite-size
 correction.  This result is different from the probability of the event that 
  only charged lepton is detected but does not contradict with that of  ordinary 
experiments,   because the boundary conditions are different in two cases.

The finite-size correction of the probability of the event that  only a charged lepton 
is detected is computed  with $S[{T}']$ which satisfies the boundary condition 
for the  charged lepton. Here   
 ${T}'={T}_{l}-{T}_{\pi}$ is the
time interval for observing the  charged lepton and the  probability 
is expressed by Eq. $(\ref{probability-3})$ with
$\omega_{\nu} \rightarrow \omega_{l}= m_{l}^2c^4/(2E_{l}\hbar)$. 
Since charged leptons  are   heavy, $\omega_{l}{T}'$ becomes 
very large  and $\tilde{g}(\omega_{l}{T}')$ becomes ${2 \over
\omega_l T} \approx 0$  for macroscopic
${T}'$.
 Thus,  the
probability  does not have a finite-size 
correction  and   agrees with that of the normal
term. 
Although the light-cone singularity forms in both cases, the 
diffraction component becomes relevant only when  the detected  particle 
is very light.

The probability of the event that  a charged
lepton is detected  depends on the boundary condition
of the neutrino. When a neutrino  is detected at ${T}_\nu$, the
charged-lepton
spectrum includes the diffraction component but, when the neutrino is not
detected,   the  charged-lepton spectrum does not include the
diffraction component. The 
latter result is standard, whereas the former is not, but may be 
verified experimentally.

We now compare   the finite-size correction  with the diffraction of light
passing  through a hole. The former is that of a quantum mechanical wave
and appears in the transition amplitude.  The 
amplitude is determined by the initial and final states, and the time interval. The 
diffraction pattern forms  in a  direction parallel to the momentum  of the 
non-stationary wave.   The size   determined by $\omega_{\nu}$ is
extremely large for light particles and stable with respect to variations 
of the energy and other
 parameters. Consequently  the diffraction is easily observed without fine tuning.
Now  the latter diffraction is that
of a classical wave and appears in its   intensity.   
The  diffraction forms  in a direction perpendicular
to the momentum and with a phase difference $\omega_\gamma^{dB} \delta t $
where $\omega_\gamma^{dB}=c|\vec{p}_\gamma|/\hbar$ of the 
stationary wave. Its shape is determined by $\omega_\gamma^{dB}$, which
is large and varies rapidly when the parameters are changed. Thus, the initial
energy must be fine tuned to observe the diffraction of light.

\section{Summary and future prospects} 

We  found  that the  finite-size correction to the probability of the
events that the neutrino from the pion decay is detected   at a
finite $T$ can be used to measure the neutrino absolute mass.
The large corrections  of    unusual properties are caused by 
the Schr\"odinger  equation and computed by 
$S[T]$ that
satisfies the boundary condition of a finite $T$.

The neutrino energy spectrum  and other patterns are determined by the   
difference of angular velocities,
$\omega_{\nu}=\omega_\nu^E-\omega_\nu^{dB}$, 
where $\omega_\nu^E={E_\nu/\hbar}$ and
$\omega_\nu^{dB}={c|\vec{p}_\nu|/\hbar}$. 
The  $\omega_{\nu} $ takes 
the  extremely small  value $  {m_{\nu}^2c^4}/{(2E_{\nu} \hbar)}$
because of the   
unique neutrino features  \cite {pdg,Tritium,WMAP-neutrino}.
Consequently, the  correction  term  becomes finite in the  macroscopic 
spatial region of $r \leq \frac{2\pi E_{\nu} \hbar c}{
m_{\nu}^2c^4}$ and affects experiments in the mass-dependent manner  at 
near-detector regions.  

 The dominant part of probability
 of the event that the electron neutrino is detected  in this region 
is the correction term and has many unusual properties. Unfortunately,
 the measurement  is not easy, because
  the identification of the
 electron neutrino and the neutrino experiment in 
this region itself are  extremely hard. 
Hence the precision experiments have not been made in this region except
 LSND so
 far, and the predicted effects are not
 in-consistent with the existing data.  This is a future problem. 

Excesses of the events are  observed by K2K \cite{excess-near-detectorK2K}, 
MiniBooNE \cite{excess-near-detectorMini,excess-near-detectorMini-n}  
and MINOS \cite{excess-near-detectorMino}, and   the excess in electron 
neutrinos known as the LSND anomaly were shown  consistent with  the 
finite-size corrections.
Because the finite-size correction are independent of flavor 
oscillations, we found the
absolute mass around  $0.082\pm 0.020$ eV, Eqs. $(\ref{masses1})$ and $(\ref{masses2})$ from 
LSND.  They   resolve the controversy between  LSND with others  \cite{LSND-ishikawa-tobita}.
We compared all previous neutrino experiments and found that the new 
contribution in near-detector
regions derived from the 
neutrino diffraction  is surprisingly consistent with all of them.
 It would be important to
confirm the neutrino diffraction and  the absolute neutrino mass 
with  precision  experiments.  
The mass $0.082\pm 0.020$ eV/{$c^2$} is close to the values suggested by  other 
experiments or observations. Recent  neutrino-less double 
$\beta$-decay experiment at KamLAND-Zen \cite{KamLAND-Zen} gave a value $0.12-0.25$ eV/{$c^2$} 
for the upper bound of the Majorana neutrino mass. Further observations 
may be able to confirm if the neutrino is Dirac or Majorana
particle. The bounds from cosmology, $0.24$  eV/{$c^2$}
\cite{pdg},   $\sum_i m_{\nu_i} \leq 0.44$ eV/{$c^2$}
\cite{WMAP-neutrino,wmap}, and $\sum_i m_{\nu_i} \leq 0.23$
eV/{$c^2$} \cite{plank},  are also close to the
present value within three flavor \cite{pdg,three-flavour}.   
 Thus the masses  Eqs. $(\ref{masses1})$, $(\ref{masses2})$, 
 $(\ref{sum-masses1})$, and $(\ref{sum-masses2})$ are consistent with existing data. 

The probability  determined by the diffraction around the overlapping 
region of wave functions of
parent and daughters  gives the finite-size correction and its magnitude 
depend on
the size of wave functions that the neutrino interact with.
We have used the values from the sizes of bound nucleus, $\sigma_b$, in  
targets 
in Eqs. $(\ref{masses1}),\ (\ref{masses2}),\ (\ref{sum-masses1}),\text{ 
and } (\ref{sum-masses2})$.
If they are extended and have larger sizes of wave functions,
 the larger values, $\sigma_m$,  are used.  In this case, the absolute 
mass becomes
\begin{eqnarray}
m_{\nu_i}=m^b_{\nu_i}({ \sigma_m \over \sigma_b})^{1/2},
\end{eqnarray}
where $m^b_{\nu_i}$ shows
the mass values 
Eqs. $(\ref{masses1}),\ (\ref{masses2}),\ (\ref{sum-masses1}),   \text{
and } (\ref{sum-masses2})$.
Since
\begin{eqnarray}
 \sigma_m \geq \sigma_b
\end{eqnarray}
the mass values 
Eqs. $(\ref{masses1}),\ (\ref{masses2}),\ (\ref{sum-masses1}),   \text{
and } (\ref{sum-masses2})$ are considered  the lower bounds. A future
precision experiment in the short distance region of detecting  the
electron neutrino may be able to test
the correction and give the precise masses.  

We described a new method derived from quantum
wave-like phenomenon  specific to  the extremely
light particle, and   showed the new physical 
observable.
The effect   of the transition amplitude  was studied in 
the lowest order  in $G_F$. Because that 
is independent of and not cancelled with higher order corrections, the 
effects due to  a propagator of $W^{\pm}$ and higher order corrections 
of the renormalized theory  do not modify the  amplitude in the lowest 
order in $W$'s mass, $M_{W}$, and the light-cone 
singularity, hence our results  are kept intact by them. 
A similar enhancement of $10^{4}$ or more caused by matter effect 
has been observed in various area \cite{positron}. Other large-scale 
quantum  phenomena  will be studied 
in subsequent presentations.

\section*{Acknowledgments}
 This work was partially supported by a 
Grant-in-Aid for Scientific Research (Grant No. 24340043). The authors  
thank Dr. Nishikawa, Dr. Kobayashi, and Dr. Maruyama for useful discussions on 
the near detector of the T2K experiment and Dr. Asai, Dr. Kobayashi,
Dr. Mori, and Dr. Yamada
for useful discussions on interferences. 

\def\thesection{Appendix \Alph{section}}
\def\thesubsection{\Alph{section}-\Roman{subsection}}
\renewcommand{\theequation}{A.\arabic{equation}}
\setcounter{equation}{0}
\setcounter{section}{0}

\section{The finite-size correction to Fermi's Golden rule}
 In computing a scattering cross section and  decay rate with  Fermi's Golden
 rules,  the $1/  T$ correction to the following  formula for a large $T$ for 
a smooth function 
$g(\omega)$,
\begin{align}
 \int d \omega g(\omega)\left({\sin \left({\omega {T}}/{2}\right) \over
 \omega}\right)^2= {T}\int dx \,g\left({x}/{{T}}\right)\left({\sin (x/2) \over
 x}\right)^2= 2\pi T g(0),
\end{align} 
 \cite{Dirac,Schiff-golden}
 diverges if 
 an   expansion,  $\displaystyle g\left({x}/{{T}}\right)=g(0)+\sum_l
 \frac{g^{(l)}(0)}{l!}\left(\frac{x}{{T}}\right)^l$ is substituted.  
 The diverging integral becomes finite with a use of boundary condition.
 Eq. (\ref{amplitude})  is such  amplitude that satisfies the boundary conditions  at  $T$, and  gives  the unique finite-size
 correction.

\section{ Light cone singularity.}

Innumerable  states at the ultra-violet energy region in a relativistic 
invariant  system lead  the correlation function  
$\Delta_{\pi,l} (\delta
 x)$  of Eq. $(\ref{pi-mucorrelation})$ to the integral form  with 
the  variable  $q=p_{l}-p_{\pi}$. Because 
\begin{eqnarray}
q^2=m_l^2+m_{\pi}^2-2p_{l}\cdot p_{\pi},
\end{eqnarray}
$q^2$ becomes negative at $|{\vec p}_l| \rightarrow \infty$, and 
the  integration  is made over the region $0 \leq q^0$ and    
$-p_{\pi}^0 \leq q^0 \leq 0$.

The integral over   $0 \leq q^0$ is expressed as, 
\begin{align}
\left[m_{\pi}^2p_{\nu}\cdot\left(p_{\pi}+i \frac{\partial}{
 \partial \delta x} \right) -2i(p_{\pi} \cdot p_{\nu})p_{\pi}\cdot\left(\frac{\partial}{
 \partial \delta x} \right)\right] \tilde I_1,  
\end{align}
where 
\begin{align}
\tilde I_1=\int d^4 q \  \frac{\theta(q^0)}{4\pi^4}\text {Im}\left[1 \over
 q^2+2p_{\pi}\!\cdot\! q+{\tilde m}^2-i\epsilon\right] e^{iq \cdot \delta x
 }\nonumber,
\end{align}
and  ${\tilde m}^2=m_{\pi}^2-m_{l}^2$. 

For ${\tilde m}^2 \geq 0$, by expanding the   integrand of $\tilde I_1$ in
$p_{\pi}\cdot q$,
we have the expression   
with the light-cone singularity
\cite{Wilson-OPE,bogoliubov-light},
 $\delta(\lambda)$,  and less singular and regular terms  that  are described with 
Bessel functions,  
\begin{align}
&\tilde I_1=2i  \left[ \frac{\epsilon(\delta t)}{4\pi}\delta(\lambda)+
 \tilde I_1^\text{regular} \right]  \label{muon-correlation-},\ 
2i 
\tilde I_1^\text{regular}=D_{\tilde m}\left(-i\frac{\partial}{\partial \delta x}\right)    f_\text{short},\nonumber\\
&f_\text{short} =
-\frac{i{\tilde m}^2}{
 8\pi \xi } \theta(-{\lambda})
\{N_1(\xi
 )-
i \epsilon(\delta t) J_1(\xi)\}
 -\frac{i{\tilde m }^2}{
 4\pi^2\xi }\theta(\lambda)K_1(\xi),\nonumber\\ 
&D_{\tilde m}\left(-i\frac{\partial}{\partial
\delta x}\right) =\sum_l
 \left(\frac{1}{l!}\right)\left(-2ip_{\pi}\!\cdot\!\left({\partial \over \partial \delta x}\right)
\frac{\partial}{\partial {\tilde m}^2}\right)^l,\ \xi=\tilde m\sqrt \lambda ,
\end{align}
where $N_1, J_1$, and $K_1$ are 
Bessel functions.

Now the region   $-p_{\pi}^0 \leq
q^0 \leq 0$ satisfies the conservation law at $T \rightarrow \infty$. The 
integral, $I_2$,  has neither  singularity nor long range part, and gives the asymptotic value. This is computed easily by integrating the coordinates first, then the standard  value is obtained. This can be computed also  numerically. 

For ${\tilde m}^2 < 0$, the expansion of the integrand in $p_{\pi}\!\cdot\! q$ is 
made with ${1 \over q^2+{\tilde m}^2 -i\epsilon}$ and the imaginary part 
vanishes in $ q^2 <0$,  and $|{\vec p}_l| \rightarrow \infty$.   
The integral from     $-p_{\pi}^0 \leq
q^0 \leq 0$, $I_2$, satisfies the conservation law which can not fulfil for 
$m_l > m_{\pi}$. Thus   $\Delta_{\pi,l} (\delta x)$ 
and   the probability to a lepton of $m_l > m_{\pi}$ vanish. 

{}

\end{document}